\documentclass{aa}
\input{epsf}					

\begin{document}

\def \etal      {et al.\ }

\def \kT  {{\rm k}T}
\def \Mg {M_{\rm gas}}
\def \Mv {M_{\rm V}}
\def \Rv {R_{\rm V}}
\def \fg {f_{\rm gas}}
\def \btm {\beta_{\rm T}}
\def \ner {n_{\rm e}(r)}
\def \ne {n_{\rm e}}
\def \nH {n_{\rm H}}

\def \MgT {\hbox{$M_{\rm gas}$--$T$} }
\def \EMT {\hbox{$EM$--$T$} }
\def \ST {\hbox{$S_{\rm I}$--$T$} } 
\def \LxT {\hbox{$L_{\rm X}$--$T$} }
\def \MvT {\hbox{$\Mv$--$T$} } \def \RvT {\hbox{$\Rv$--$T$} } \def
\betamodel {\hbox{$\beta$--model} } \def \xc {x_{\rm c}}

\def \rhocrit {\hbox{\rho_c}}
\def \Del  {\Delta_{\rm c}(\Omz,\Lambda)}
\def \ho  {{\rm H_{0}}}
\def \qo  {{\rm  q_{0}}}
\def \Omo {\rm \Omega_{0}}
\def \Omz {\rm \Omega_{\rm z}}
\def \hc {\rm h_{50}} 
\def \keV {\rm keV} \def \msun
{\hbox{$M_{\sun}$}} \def \k {{\rm k}} \def \G {{\rm G}} \def \mp {{\rm
m_{p}}}

\def \eps {{\epsilon}}
\def \Dzf {(\Del \Omo)/(18\pi^{2} \Omz)}
\def \Dz {\Delta_{\rm z}}
\def \dA {d_{\rm A}(z)}

\title{The X-ray surface brightness profiles of hot galaxy
clusters up to $z\sim 0.8$: evidence for self-similarity and
constraints on $\Omo$} \author{ M. Arnaud\inst{1}, N. Aghanim\inst{2},
D.M. Neumann\inst{1}} \institute{ $^1$ C.E.A., DSM, DAPNIA, Service
d'Astrophysique, C.E. Saclay, F-91191, Gif-Sur-Yvette Cedex, France \\
$^2$Institut d'Astrophysique Spatiale, Universit\'e Paris-Sud, F-91405
Orsay Cedex, France\\
email: marnaud@discovery.saclay.cea.fr; aghanim@ias.fr; ddon@cea.fr}

\offprints{M. Arnaud, marnaud@discovery.saclay.cea.fr}

  \date{Received 19 October 2001 / Accepted 7 March 2002}

\titlerunning{The X-ray surface brightness profiles of hot galaxy
clusters  up to $z\sim 0.8$} \authorrunning{Arnaud \etal}

\abstract{We study the surface brightness profiles of a sample of 25
distant $(0.3<z<0.83)$ hot $(\kT > 3.5~\keV)$ clusters, observed with
ROSAT, with published temperatures from ASCA. For both open and flat
cosmological models, the derived emission measure profiles are scaled
according to the self-similar model of cluster formation.  We use the
standard scaling relations of cluster properties with redshift and
temperature, with the empirical slope of the \MgT relation derived by
Neumann \& Arnaud (\cite{neumann01}).  Using a $\chi^2$ test, we
perform a quantitative comparison of the scaled emission measure
profiles of distant clusters with a local reference profile derived
from the sample of 15 hot nearby clusters compiled by Neumann \&
Arnaud (1999), which were found to obey self-similarity.  This
comparison allows us to both check the validity of the self-similar
model across the redshift range $0.04-0.8$, and to constrain the
cosmological parameters.  \\
For a low density flat universe, the scaled distant cluster data were
found to be consistent, both in shape and normalisation, with the
local reference profile.  It indicates that hot clusters constitute a
homologous family up to high redshifts, and gives support to the
standard picture of structure formation for the dark matter component. 
Because of the intrinsic regularity in the hot cluster population, the
scaled profiles can be used as distance indicators, the correct
cosmology being the one for which the various profiles at different
redshifts coincide.   The intrinsic limitations of the method, in
particular possible systematic errors and biases related to the model
uncertainties, are discussed.  Using the standard evolution model,
the present data allow us to put a tight constraint on $\Omo$ for a
flat Universe: $\Omo=0.40^{+0.15}_{-0.12}$ at 90\% confidence level
 (statistical errors only).  The critical model ($\Omo=1$) was
excluded at the 98\% confidence level.  Consistently, the observed
evolution of the normalisation of the \LxT relation was found to
comply with the self-similar model for $\Omo=0.4$, $\Lambda=0.6$.  The
constraint derived on $\Omo$ is in remarkable agreement with the
constraint obtained from luminosity distances to SNI or from combined
analysis of the power spectrum of the 2dF galaxy redshift Survey and
the Cosmic Microwave Background anisotropies.  }
 
\maketitle

\section{Introduction}

In the simplest models of structure formation, purely based on
gravitation, galaxy clusters constitute a homologous family.  Clusters
are self-similar in shape, and predictable scaling laws relate each
physical property to the cluster total mass $M$ and redshift $z$
(Kaiser \cite{kaiser86}; Navarro \etal \cite{navarro97}; Teyssier \etal
\cite{teyssier}; Eke \etal \cite{eke}; Bryan \& Norman
\cite{bryan98}).  Self-similarity applies to both the dark matter
component and the hot X--ray emitting intra-cluster medium (ICM).

From the observation of the ICM, we do see regularity in the local ($z
<0.1$) population of clusters, like strong correlations between
luminosity, gas mass, total mass, size and temperature $T$ (Mohr \etal
\cite{mohr97}; Allen \& Fabian \cite{allen}; Markevitch
\cite{markevitch}; Arnaud \& Evrard \cite{arnaud}; Horner \etal
\cite{horner}; Mohr \etal \cite{mohr99};Vikhlinin, \etal
\cite{vikhlinin99}; Nevalainen \etal \cite{nevalainen}; Finoguenov
\etal \cite{finoguenov}; Xu \etal \cite{xu}).  Furthermore, there is
strong indication of a universal shape for the density and temperature
profiles of hot ($\kT > 4~\keV$) clusters, beyond the cooling flow
region (Markevitch \etal \cite{markevitch98}; Neumann \& Arnaud
\cite{neumann99}, \cite{neumann01}; Vikhlinin, \etal
\cite{vikhlinin99}; Irwin \& Bregman \cite{irwin}; Arnaud
\cite{arnaud01}).  However, clusters also deviate from the simplest
self-similar model.  The most remarkable deviation is the slope of the
luminosity--temperature (\LxT) relation, which is steeper than
predicted.  In a recent study, Neumann \& Arnaud (\cite{neumann01})
showed that a steepening of the \MgT relation ($\Mg \propto T^{1.94}$,
instead of the standard relation $\Mg \propto T^{1.5}$) can explain the
observed \LxT relation in the hot temperature domain, and account for
the scaling properties of the normalisation of the emission measure
profiles of hot clusters.  Similar steepening was derived from direct
studies of the \MgT relation independently carried out (Mohr \etal
\cite{mohr99}; Vikhlinin, \etal \cite{vikhlinin99}) and is also
consistent with the observed slope of the isophotal size--temperature
(\ST) relation (Mohr \etal \cite{mohr97}).

Several physical processes have been suggested to explain the
departure from the simplest self-similar model.  Pre-heating by early
galactic winds, has been proposed to explain the steepening of the
\LxT relation (e.g. Kaiser \cite{kaiser91}; Evrard \& Henry \cite{evrard91}),
although other effects like AGN heating (e.g. Valageas \& Silk
\cite{valageas}; Wu \etal~\cite{wu}), radiative cooling (Pearce
\etal~\cite{pearce}; Muawong \etal \cite{muawong}) or variation of the galaxy
formation efficiency with system mass (Bryan \cite{bryan00}) might
also play a role.  Further evidence of the importance of
non-gravitational processes is provided by the excess of entropy (the
``entropy floor'') in poor clusters (Ponman \etal \cite {ponman};
Lloyd-Davis \etal \cite{lloyd}).  Recent numerical simulations (Bialek
\etal \cite{bialek}) including pre-heating, with an initial entropy
level consistent with this observed entropy floor, do predict a
steepening of the \LxT, \MgT and \ST relations, consistent with the
observations quoted above (see also Loewenstein~\cite{loewenstein};
Tozzi \& Norman \cite{tozzi}; Brighenti \& Mathews~\cite{brighenti};
Borgani \etal~\cite{borgani}).  However, it is unclear if such a
scenario is also consistent with the level of self-similarity in shape
observed in hot clusters.  Although it is predicted that cool clusters
should have a more extended atmosphere than hot clusters (e.g. Tozzi
\& Norman \cite{tozzi}), to our knowledge no detailed study on the
relationship between internal shape and cluster temperature,
specifically for relatively hot clusters, has been carried out so far.

The evolution of cluster X--ray properties is an essential piece of
information to reconstruct the physics of the formation processes for
the gas component and can also be used as a cosmological test.  Models
with pre-heating predict an absence of evolution in the \LxT and \MgT
relations, at least up to $z\sim 0.5$ (e.g. Bialek \etal
\cite{bialek}).  There is some indication, based on a few massive
clusters, that the $\LxT$ relation is evolving weakly, if at all
(Mushotzky \& Scharf \cite{mushotzky}; Sadat \etal \cite{sadat};
Donahue \etal \cite{donahue}; Reichart \etal \cite{reichart};
Schindler \cite{schindler99}; Fairley \etal \cite{fairley}).  Several
groups (Sadat \etal \cite{sadat}; Reichart \etal \cite{reichart};
Fairley \etal \cite{fairley}) quantified the evolution of the
normalisation of the $\LxT$ relation, assuming it varies as
$(1+z)^\eta$.  For a critical density Universe, they found $\eta$
values significantly smaller than the theoretical prediction
$\eta=1.5$ in the self-similar model and consistent with no-evolution. 
However, the luminosity estimates depend on the assumed cosmological
parameters and so does the constraint on the evolution parameter
(Fairley \etal \cite{fairley}; Reichart \etal \cite{reichart}).  The
evolution of other scaling laws like the gas or total mass temperature
relation are even more poorly known (Schindler~\cite{schindler99};
Matsumoto \etal \cite{matsumoto}).

Using non-evolving physical properties of clusters as distance
indicators can provide interesting constraints on cosmological
parameters, such as the density parameter, $\Omo$, and the
cosmological constant, $\Lambda$.  In this context, the gas mass
fraction has been considered by Pen (\cite{pen}), although present
constrains are poor (Rines \etal \cite{rines}; Ettori \& Fabian
\cite{ettori}).  Recently, Mohr \etal (\cite{mohr00}) measured the \ST
relation for a sample of intermediate redshift clusters, $0.2 < z
<0.55$.  Using standard cluster evolution models, they argue that this
relation should not evolve with redshift.  They did find that the
intermediate redshift data are consistent with the local relation and
were able to rule out a critical density Universe.

With the present study, we aim at a better understanding of the
evolution of the scaling and structural properties of hot clusters
with redshift.  Furthermore, we show that strong constraints on the
cosmological parameters can be drawn, based on the cluster scaling
properties.

We perform for the first time a systematic study of the X-ray surface
brightness profiles of distant ($0.3 <z < 0.83$) hot ($\kT > 3.5
\keV$) clusters, measured with the ROSAT satellite.  This sample is
combined to the sample of local ($z\sim 0.05$) clusters, presented in
Neumann \& Arnaud (\cite{neumann99}).  The surface brightness profile
is directly related to the emission measure profile (or equivalently
to the gas density profile).  Comparing the profiles of clusters at
different redshifts and temperatures obviously provides more
information than simply considering global quantities such as the
total X-ray luminosity or punctual quantities like the isophotal
radius.  With the present study, we wish to address the following
issues i) Do hot clusters remain self-similar in shape up to high
redshift?  ii) How do the scaling properties of the profiles with
redshift compare quantitatively with the theoretical expectations of
the self-similar model?  iii) What constraints can we put on the
cosmological parameters from these data?  iv) Is the evolution of the
\LxT relation really inconsistent with a self-similar model?

The paper is organized as follows.  In Section 2, we present the
cluster sample and the data analysis performed to derive the surface
brightness profiles and then the emission measure profiles.  In
Section 3 we derive how the emission measure profiles should scale
with redshift, depending on cosmological parameters, for the
self-similar model of cluster formation.  In Section 4, we derive the
corresponding scaled emission measure profiles for our cluster sample,
that we use, in Section 5, to test the self-similar model and constrain
the cosmological parameters.  In Section 6, we study the 
the \LxT relation.  In Section 7 we discuss our results and Section 8
contains our conclusions.

The present
time Hubble constant in units of 50 km/s/Mpc is noted $\hc$ in the
following. The data analysis is done with $\hc=1$.

\begin{table*}
\caption[]{Basic data for the 25 distant clusters in the sample.    }
\begin{center}
\begin{tabular}{l|lccc|rccccc}
\hline
Cluster & $z^{\rm a}$ & $\kT$ & $L_{\rm bol}^{\rm ASCA}$ & Ref$^{\rm
b}$ & $t_{\rm exp}$$^{\rm c}$& $CR_{\rm det}$&$R_{\rm
det}$ &Emiss.&$L_{bol}^{\rm Rosat}$ \\
      &     &  (keV) &  ($10^{45}$ erg/s) &&(ksec)&
      ($10^{-2}$ ct/s) &(') & ${\rm (ct/s/10^{13}cm^{-5})}$& ($10^{45}$ erg/s)\\
\hline
ACCG 118         &0.308&  $12.1^{+0.9}_{-0.5}$&5.52&6&14~(p) 
&$25.2\pm0.5$&5.9&4.88&$7.1\pm0.1$\\
CLG 0016+1609    &0.541&  $8.9^{+0.6}_{-0.6}$&7.29&4&43~(p) 
&$7.8\pm0.1$&4.5&5.21&$5.6\pm0.1$\\
A 370            &0.373&  $6.6^{+0.6}_{-0.5}$&1.75&7&32~(h) 
&$2.5\pm0.1$&2.0&2.15&$2.3\pm0.1$\\
CL 0302.7+1658   &0.424&  $4.4^{+0.8}_{-0.6}$&1.08&4&34~(h) 
&$0.72\pm0.07$&1.0&1.64&$1.0\pm0.1$\\
MS 0353.6-3642   &0.32&  $6.5^{+1.0}_{-0.8}$&1.43&4&22~(h) 
&$3.3\pm0.2$&2.7&2.45&$1.7\pm.0.1$\\
MS 0451.6-0305   &0.55&  $10.3^{+0.9}_{-0.8}$&6.71&4&16~(p) 
&$7.3\pm0.2$&4.5&4.98&$6.1\pm0.2$\\
MS 0811.6+6301   &0.312&  $4.9^{+1.0}_{-0.6}$&0.570&4&147~(h) 
&$0.96\pm0.05$&1.3&2.02&$0.58\pm0.03$\\
MS 1008.1-1224   &0.301&  $8.2^{+1.2}_{-1.1}$&1.84&4&69~(h) 
&$2.6\pm0.1$&2.3&1.72&$2.0\pm0.1$\\
A 959            &0.353&  $7.0^{+1.1}_{-0.8}$&2.30&6&16~(p) 
&$7.2\pm0.2$&5.5&5.40&$1.81\pm0.06$\\
MS 1054.4-0321   &0.83&  $10.5^{+2.1}_{-1.3}$&4.39&6&191~(h) 
&$0.9\pm0.2$&2.0&2.30&$4.9\pm1.0$\\
A 1300           &0.3058&  $11.4^{+0.8}_{-0.6}$&6.73&8&8.6~(p) 
&$18.2\pm0.5$&5.5&4.56&$5.2\pm0.1$\\
MS 1137.5+6625   &0.782&  $5.7^{+1.3}_{-0.7}$&1.62&2&99~(h) 
&$0.64\pm0.05$&1.0&2.80&$2.0\pm0.1$\\
MS 1224.7+2007   &0.327&  $4.1^{+0.7}_{-0.5}$&0.690&4&41~(h) 
&$0.82\pm0.07$&1.0&2.19&$0.58\pm0.05$\\
MS 1241.5+1710   &0.54&  $6.1^{+1.4}_{-1.1}$&2.26&4&31~(h) 
&$1.3\pm0.1$&1.2&2.48&$2.0\pm0.2$\\
A 1722           &0.3275&  $5.9^{+0.3}_{-0.3}$&1.77&6&28~(h) 
&$3.0\pm0.2$&2.3&2.41&$1.5\pm0.1$\\
RX J1347.5-1145  &0.451&  $9.3^{+0.7}_{-0.6}$&21.0&9&36~(h) 
&$12.7\pm0.3$&3.0&1.96&$17.4\pm0.4$\\
Zwcl 1358+6245   &0.328&  $6.9^{+0.5}_{-0.5}$&2.14&4&23~(p) 
&$8.0\pm0.2$&2.7&5.20&$1.95\pm0.05$\\
3C295            &0.46&  $7.1^{+1.3}_{-0.8}$&1.90&6&29~(h) 
&$2.1\pm0.1$&1.0&2.52&$2.9\pm0.1$\\
MS 1426.4+0158   &0.32&  $6.4^{+1.0}_{-1.2}$&0.970&4&37~(h) 
&$1.4\pm0.1$&1.3&2.16&$1.09\pm0.08$\\
A 1995           &0.318&  $10.7^{+1.5}_{-1.1}$&2.82&6&38~(h) 
&$4.3\pm0.2$&2.0&2.21&$3.7\pm0.1$\\
MS 1512.4+3647   &0.372&  $3.4^{+0.4}_{-0.4}$&0.920&4&35~(h) 
&$1.8\pm0.1$&2.0&2.52&$0.86\pm0.07$\\
MS 1621.5+2640   &0.426&  $6.6^{+0.9}_{-0.8}$&1.58&4&44~(h) 
&$1.06\pm0.09$&1.3&2.13&$1.7\pm0.1$\\
RX J1716.4+6708  &0.813&  $5.7^{+1.4}_{-0.6}$&1.15&3&122~(h) 
&$0.37\pm0.03$&1.0&2.34&$1.5\pm0.1$\\
MS 2137.3-2353   &0.313&  $4.9^{+0.3}_{-0.3}$&3.35&4&10~(p) 
&$16.2\pm0.4$&3.5&5.06&$2.91\pm0.07$\\
ACCG 114         &0.312&  $9.8^{+0.6}_{-0.5}$&3.25&1&23~(h) 
&$6.5\pm0.3$&3.8&2.26&$4.1\pm0.2$\\
\hline
\end{tabular}
\end{center}
\smallskip Notes: The values of the bolometric luminosities, $L_{\rm
bol}$, are for $\Omo = 1$ ($\qo=0.5$) and $\ho = 50$kms/s/Mpc.  All
errors are at the $68\%$ confidence level.

\noindent
$^{a}$  The redshifts, $z$, are taken from NED.

\noindent $^{b}$ References for the temperature and ASCA luminosities
listed column (3) and (4): 1.  Allen \& Fabian \cite{allen}; 2.
Donahue \etal \cite{donahue}; 3.  Gioia \etal \cite{gioia}; 4.  Henry
\cite{henry}; 5.  Jeltema \etal \cite{jeltema}; 6.  Mushotzky \&
Scharf \cite{mushotzky}; 7.  Ota \etal \cite{ota}; 8.  Pierre \etal
\cite{pierre}; 9.  Schindler.  \etal \cite{schindler97}.  The
temperature errors published at the $90\%$ confidence level were
divided by 1.65 to estimate the $68\%$ confidence level errors.
Luminosities published in the $[2-10] \keV$ energy band (references 4.
and 7.)  were converted to bolometric luminosity using a MEKAL model
with the temperature given column (3).  When necessary, the published
luminosities were corrected for $\ho = 50$kms/s/Mpc and $\qo=0.5$
(references 2., 3.  and 6.).

\noindent
$^{c}$  The letter in parenthesis stands for the ROSAT detector used,
(h) for HRI, (p) for PSPC.

\label{tab:data}
\end{table*}

\section{The data}

\subsection{The cluster sample}
We considered all distant ($z>0.3$) clusters observed by ROSAT, with
published ASCA temperatures.  We believe our original list was
complete with respect to ROSAT public archival data and publications,
available at the end of 1999.  We excluded three clusters with no
obvious X-ray center: the double cluster A851 (Schindler \etal
\cite{schindler98}), the clumpy cluster Cl 0500-24 (Schindler \&
Wambsganss \cite{schinwamb97}) and MS 1147.3+1103 (the HRI image shows
a very flat elliptical morphology in the core, with some evidence of
bimodality).  The derivation of a surface brightness profile for those
clusters would have been arbitrary.  We also excluded Cl 2244-0221 and MG
2053.7-0449 (Hattori \etal \cite{hattori}) due to the too poor
statistical quality of the HRI data.

The list of the 25 distant clusters selected is shown in
Tab.\ref{tab:data}, as well as the exposure times and the ROSAT
detector used.  The sample covers a redshift range of $z=0.3-0.83$. 
We also give in the table the temperatures and bolometric
luminosities, measured with ASCA. The only exception is MS1054, for
which we list the recent Chandra temperature estimate of the main
cluster component (Jeltema \etal \cite{jeltema}), the western
subcluster (see Neumann \& Arnaud \cite{neumann00}) being excluded in
our spatial analysis below.  When several temperature estimates for a
given cluster were published, we have chosen the most recent analysis using
the latest ASCA calibrations.  The various published values were
usually consistent.

\begin{figure*}[t]
\epsfxsize=17.5cm \epsfbox{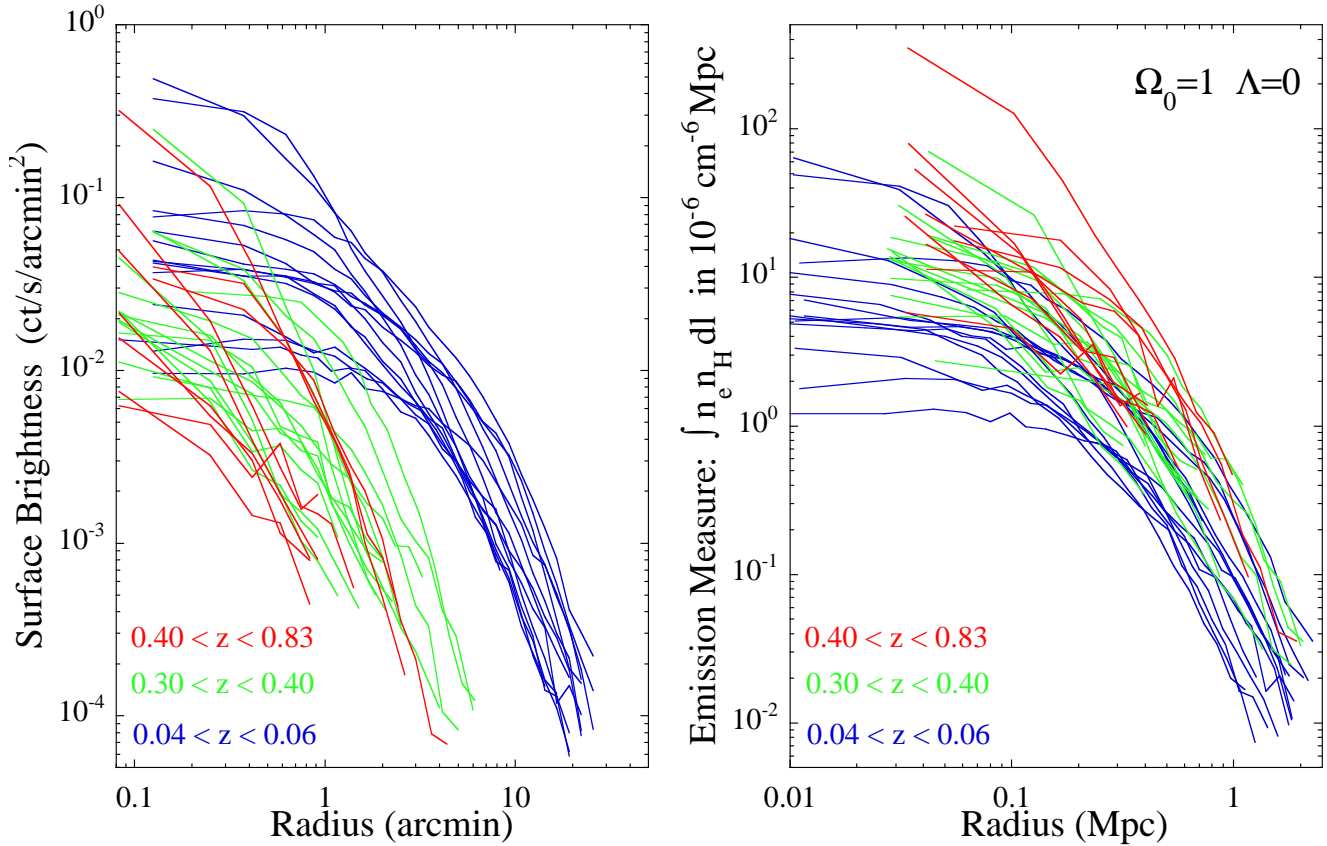} \caption{ The ROSAT X-ray surface
brightness ($S_{\rm X}$) profiles (left panel) and emission measure
($EM$) along the line of sight profiles (right panel) of all (nearby
and distant) clusters in the sample.  A straight line is drawn between
the data points for each cluster for a better visualization of the
profiles.  The curves are color-coded by redshift.  The $S_{\rm X}$
profiles are background subtracted and corrected for vignetting
effects.  $EM$ is derived from the surface brightness taking into
account the $(1+z)^{-4}$ cosmological dimming and the emissivity in
the ROSAT band (see Eq.~\ref{eq:sx2em}).  The angular radius is
converted to physical radius, assuming $\Omo=1$ and $\hc=1$.  }
\label{fig:sxem}
\end{figure*}

To study cluster evolution, we combined this new distant cluster
sample with the sample considered in our previous study of the surface
brightness profiles of nearby clusters (Neumann \& Arnaud
\cite{neumann99},\cite{neumann01}).  This nearby cluster sample
comprises 15 Abell clusters in the redshift range $0.04<z<0.06$, which
were observed in pointing mode with the ROSAT PSPC with a high signal
to noise ratio and for which accurate temperature measurements exist
from the literature (see Neumann \& Arnaud \cite{neumann99} for
details).  

We emphasize that the study presented here focuses on relatively hot
clusters, the minimum temperature for the nearby and distant cluster
samples being 3.7 and 3.4 $\keV$ respectively.

\subsection{Surface brightness profiles}

The surface brightness profile of each cluster, $S(\theta)$, was
constructed using the standard procedures described in Neumann \&
Arnaud (\cite{neumann99}).  We only considered photons in the energy band
0.5-2.0~keV for the PSPC data and only took into account channels 2-10
for the HRI data, in order to optimize the signal-to-noise (S/N) ratio.  We
binned the photons into concentric annuli centered on the maximum of
the X-ray emission with a width of 15" and 10" per annulus for the
PSPC and HRI data respectively.  We cut out serendipitous sources in
the field of view or cluster substructures, if they show up as a local
maximum.  The HRI particle background was subtracted from the HRI
profiles using the background map constructed for each observation
with the method of Snowden \etal (\cite{snowden}).  The vignetting
correction was performed using the exposure maps computed with {\it
EXSAS} (Zimmermann \etal \cite{zimmermann}) for the PSPC data and with
the software developed by Snowden (\cite{snowden}) for the HRI data.
The X-ray background for each pointing was estimated using vignetting
corrected data in the outer part of the field of view and subtracted
from the profile.  A $5\%$($10\%$) systematic error was added
quadratically to the statistical error on the PSPC(HRI) background
level.

For 5 clusters both HRI and PSPC data were available.  We found an
excellent agreement between the HRI and PSPC profiles, except for the
most inner radial bin, where the effect of the wider PSPC/PSF can be
observed.  This blurring is clearly negligible at larger radii.  If
available, we thus always choose the PSPC data, due to its higher
intrinsic sensitivity and lower background level, which allow to trace
the cluster emission further out.

To avoid too noisy profiles, we rebinned the data, for both the nearby
and distant cluster samples, so that the variations of $S(\theta)$
from bin to bin are significantly larger than the corresponding
statistical error and thus representative of the cluster shape. 
Starting form the central annulus, we regrouped the data in adjacent
annuli so that i) at least a S/N ratio of $3\sigma$ is reached after
background subtraction and ii) the width of the annulus,
$\Delta(\theta)$, at radius $\theta$, has a size at least
$0.15~\theta$.  This logarithmic binning insures a roughly constant
S/N ratio for each bin in the outer part of the profiles, where the
background can still be neglected (the S/N ratio would be constant for
a $\betamodel$ with $\beta=2/3$ and no background).  The adopted
rebinning was found to be a good compromise between the desired
accuracy and a reasonable sampling of the profiles.

The resulting surface brightness profiles are shown in
Fig.~\ref{fig:sxem} (left panel) for the distant and nearby clusters. 
For each cluster the data points are connected by a straight line to
guide the eye.  The profiles are plotted up to the adopted detection
limit of $3\sigma$ above background.  The corresponding detection
radius, as well as the total ROSAT count rate, $CR_{\rm det}$, within
this radius is given in Tab.~\ref{tab:data} for each cluster.  In
Fig.~\ref{fig:sxem} and Fig.~\ref{fig:emsc}, we color--coded the
profiles according to cluster redshift: blue for nearby clusters
($0.04 < z < 0.06$), green for moderately distant clusters ($0.3 < z
<0.40$) and red for very distant clusters ($0.40 < z < 0.83$).  Note
that this redshift subsampling of the distant cluster sample is for
display only and, unless explicitly stated, is not used in the
statistical analysis below.

\subsection{Emission measure profiles}
The emission measure along the line of sight at radius $r$,
$EM(r)=\int\ne\nH~dl$, can be deduced from the X-ray surface
brightness, $S(\theta)$:
\begin{eqnarray}
EM(r) &=& \frac{4~\pi~(1+z)^{4}~S(\theta)}{\epsilon(T,z)} \nonumber\\
&=& 4.81~10^{-7} (1+z)^{4} \left(\frac{S(\theta)}{\rm
ct/s/arcmin^{2}}\right) \nonumber\\
&&\times \left(\frac{\epsilon(T,z)}{\rm
ct/s/10^{10}cm^{-5}}\right)^{-1}~{\rm cm^{-6}~Mpc}
\label{eq:sx2em}\\
r &=&
d_{\rm A}(z)~\theta 
\label{eq:teta2r}
\end{eqnarray}
where $d_{\rm A}(z)$ is the angular distance at redshift $z$.

$\eps(T,z)$ is the emissivity in the considered ROSAT band,
$[E_{1}-E_{2}]$, taking into account the interstellar absorption and
the instrumental spectral response:
\begin{equation}
\eps(T,z) =\int_{E_{1}}^{E_{2}} S(E)e^{-\sigma(E)N_{\rm
H}}f_{T}((1+z)E)(1+z)^{2}dE
\end{equation}
where $S(E)$ is the detector effective area at energy $E$, $\sigma(E)$
the absorption cross section, $N_{\rm H}$ the hydrogen column density
along the line of sight, $f_{T}((1+z)E)$ the emissivity in photons
cm$^{3}$/s/keV at energy $(1+z)E$ for a plasma of temperature $T$.  It
was computed for each cluster using a redshifted thermal emission
model (Mewe \etal \cite{mewe1},\cite{mewe2}; Kaastra \cite{kaastra};
Liedahl \etal \cite{liedahl}), the ROSAT response (Zimmermann \etal
\cite{zimmermann}) and the $N_{\rm H}$ value estimated with the w3nh
tools available at HEARSAC (Dickey \& Lockman \cite{dickey}).  The
emissivity, $\eps(T,z)$, depends weakly on cluster temperature and
redshift in the ROSAT band (Tab.~\ref{tab:data}).

The derived $EM(r)$ profiles are shown in the right
panel of Fig.~\ref{fig:sxem} for a critical density Universe ($\Omo =1,
\Lambda=0.)$.  As will be discussed later, one can already note that
the distant clusters appear brighter than the nearby clusters.

\section{Theoretical scaling laws}

In our derivation of the theoretical emission measure profiles, as a
function of redshift and cosmological parameters, we will consider
both a flat Universe ($\Omo + \Lambda=1$) and an open Universe
($\Omo<1,\Lambda=0$).  The matter density parameter at redshift $z$ is
noted $\Omz$; $\Omz = \Omo (1+z)^{3}/E^{2}(z)$, where $E^{2}(z)
=\Omo(1+z)^{3} + (1-\Omo -\Lambda)(1+z)^{2} +\Lambda$.

\subsection{The self-similar model}

The simplest self-similar model (e.g. Bryan \& Norman \cite{bryan98}; Eke \etal
\cite{eke}) assumes that i) at a given redshift the relaxed virialized
portion of clusters corresponds to a fixed density contrast as
compared to the critical density of the Universe at that redshift ii)
the internal structure of clusters of different mass and $z$ are
similar.

The virial mass $\Mv$ and radius $\Rv$ then scale with
redshift and temperature via the well known relations:
\begin{eqnarray}
\Mv & = & 2.835~10^{15}~\btm^{3/2}\
\Dz^{-1/2}\ (1+z)^{-3/2} \nonumber \\
&& \times \left(\frac{\kT}{10\ \keV}\right)^{3/2}\
\hc^{-1}~~\msun
\label{eq:mv}\\
\Rv & = & 3.80~\btm^{1/2}\
\Dz^{-1/2}\ (1+z)^{-3/2}\nonumber \\
&& \times \left(\frac{\kT}{10\ \keV}\right)^{1/2}\
\hc^{-1}~~{\rm Mpc}
\label{eq:rv} \\
{\rm with}\nonumber \\
\Dz & = & \Dzf
\label{eq:dz}
\end{eqnarray} where $\Del$ is the density contrast (a function of
$\Omo$ and $\Lambda$) and $\btm$ is the normalisation of the virial
relation, $\G\Mv/2\Rv= \btm \kT$.

The \MvT and \RvT relations depend on the cosmological parameters
through the factor $\Del \Omo/ \Omz$.  This factor is constant with
redshift and equal to $18\pi^{2}$ for a critical density Universe. 
Analytical approximations of $\Del$, derived from the top-hat
spherical collapse model assuming that clusters have just virialized,
are given in Bryan \& Norman (\cite{bryan98}):
\begin{eqnarray}
    \Del &=&18\pi^{2} + 60w -
     32w^{2}~{\rm for}~\Omo<1,\Lambda=0  \nonumber \\
     \Del &=&18\pi^{2} + 82w -
     39w^{2}~{\rm for}~\Omo+\Lambda=1 \nonumber \\
{\rm with}~~w & = & \Omz-1
\label{eq:dc}
\end{eqnarray}
As we consider lower and lower values of the density parameter $\Omo$,
the assumption of recent cluster formation is less and less valid and
in principle, the difference between the observing time and the time
of collapse has to be taken into account (e.g. Voit \& Donahue
\cite{voit}).   If the effective formation epoch of a cluster of a
given mass is earlier than the observing time, when the Universe was
denser, the actual cluster temperature is underestimated by the recent
formation approximation.  This effect is expected to increase with
decreasing $\Omo$ and mass of the system.  Estimating accurately the
impact on the \MvT relation (mean relation and scatter) is however not
trivial, because it requires a precise modelling of cluster formation
history, the growth of clusters by continuous accretion and merger
events, and the complex physics of the ICM (Voit~\cite{voit00},
Afshordi \& Cen~\cite{afshordi}).  However, recent analysis of the
\MvT relation derived from numerical simulations suggest that the
effect of formation redshift is negligible, at least when considering
measured X--ray temperatures (Mathiesen~\cite{mathiesen}).  We will
thus neglect this effect here and use the above equations estimated at
the observed cluster redshift.

The constant $\btm$ depends on the cluster internal structure.  Its
value can be determined from numerical simulations.  The various
results agree within typically  $\pm 20 \%$, with no obvious dependence on
cosmological parameters (Henry \cite{henry}).  As in our previous work
(Neumann \& Arnaud~\cite{neumann99};\cite{neumann01}), we will adopt
the normalisation of Evrard \etal (\cite{evrard}), $\btm=1.05$.  Note
that our results do not depend on the exact value of $\btm$.

\subsection{The theoretical emission measure profiles}
The central emission measure along the line of sight
is related to the electron density profile of the gas, $\ner$, via:
\begin{eqnarray}
EM_{0} & = &2~(\nH/\ne)
\int_{0}^{\Rv} \ner^{2}~dr
\label{eq:em0}\\
\end{eqnarray}
whereas the  gas mass is given by:
\begin{eqnarray}
\Mg & = &
\mu'~\mp~(\nH/\ne)~4\pi~\int_{0}^{\Rv} \ner~r^{2}~dr
\label{eq:mg}\end{eqnarray}
where $\mp$ is the proton mass, $\mu'=1.347$ and $\nH/\ne=0.852$, for an
ionized plasma with a metallicity of 0.3 solar value.  In self-similar
models, which we consider here, the density profile can be written:
\begin{eqnarray}
\ner& =&\ne(0)\ f_{\rm n}(x)~~~;x=r/\Rv
\label{eq:ne}
\end{eqnarray}
where $x$ is the radius scaled to the virial radius and $f_{\rm n}$ is
a universal function, the same for all clusters.  By combining the
above equations, $EM_{0}$ varies as $EM_{0} \propto Q_{\rm n}
\Mg^{2}/\Rv^{5}$, where we have introduced a constant form factor
$Q_{\rm n}$, which only depends on the cluster's `universal' shape:
\begin{eqnarray}
Q_{\rm n} &= & \frac{\int_{0}^{1} f_{\rm n}^{2}(x)~dx}{9\ \left(\int_{0}^{1}
f_{\rm n}(x)~x^{2}~dx\right)^{2}}
\label{eq:q}
\end{eqnarray}
Assuming a standard $\betamodel$ with $\beta=2/3$ and a scaled core
radius of $\xc=0.123$, which fits well the scaled profiles of nearby
clusters (Neumann \& Arnaud \cite{neumann99}), gives $Q_{\rm n}=69.4$.

The scaling law for the central emission measure can now be derived
from Eq.~\ref{eq:em0},\ref{eq:mg},\ref{eq:ne} and \ref{eq:q} and the
\MvT and \RvT relations (Eq.~\ref{eq:mv},\ref{eq:rv}), assuming that
all clusters have the same gas mass fraction $\fg$:
\begin{eqnarray}
EM_{0}& = & 4.1~10^{-6}\ \left(\frac{\btm}{1.05}\right)^{1/2}\
\left(\frac{\fg}{0.2}\right)^{2}\
\left(\frac{Q_{\rm n}}{69.4}\right)\nonumber \\
&&\times~\Dz^{3/2}\ (1+z)^{9/2}\ \left(\frac{\kT}{10\ \keV}\right)^{1/2}
\hc^{3}\ {\rm cm^{-6}
Mpc}
\label{eq:em0sc}
\end{eqnarray}
This assumes that the gas mass scales as the total mass, i.e $\Mg
\propto T^{1.5}$.  The corresponding emission measure profiles can
thus be written:
\begin{eqnarray}
EM(r)& = & EM_{0}\ f_{\rm EM} (r/\Rv)
\end{eqnarray}
where $f_{\rm EM}(x)$ is the dimensionless function:
\begin{eqnarray}
f_{\rm EM}(x)& = &\int_{x}^{1} \frac{f_{\rm n}(u)}{\sqrt{u^{2}-x^{2}}}
d(u^{2})
\end{eqnarray}

As can be seen from Eq.~\ref{eq:rv} and Eq.~\ref{eq:em0sc}, the virial
radius decreases with redshift while the central emission measure
increases.  Clusters of a given mass are denser at high redshift,
following the evolution of the Universe mean density.  We thus expect
that clusters of given temperature appear smaller and brighter with
increasing redshift.

\begin{figure*}
\epsfysize=20.cm \center \epsfbox{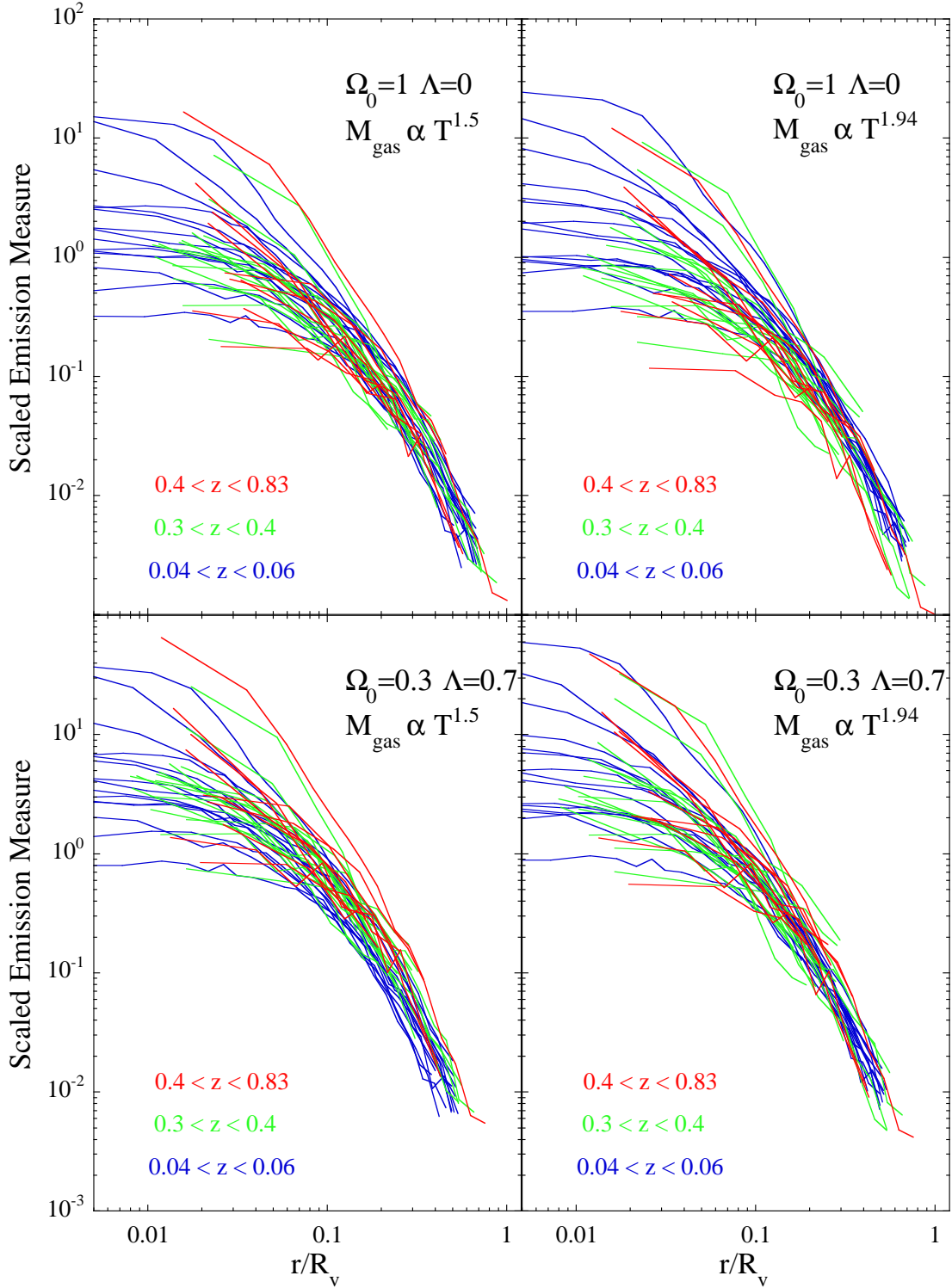} \caption{The scaled
emission measure along the line of sight as a function of scaled
radius for the 40 clusters of the sample in two cosmological models. 
Top panels: $\Omo=1., \Lambda=0$.  Bottom panels: $\Omo=0.3,
\Lambda=0.7$.  The data points for each cluster are connected by a
straight line for a better visualization of the profiles.  The curves
are color-coded by redshift.  The radius is scaled by the virial
radius, $\Rv$, computed using the theoretical scaling law
(Eq.~\ref{eq:rv}) with the measured temperature and redshift of the
clusters.  The emission measure profiles in the left panels have been
scaled by $(\Del \Omo/\Omz)^{3/2} (1+z)^{9/2} T^{1/2}$, according to
the standard self-similar model (Eq.~\ref{eq:em2emsc}), with $\Mg
\propto T^{1.5}$.  Note the remarkable similarity of the profiles at
radii larger than $\sim 0.1 \Rv$.  The scaling procedure has
significantly decreased the dispersion between the emission measure
profiles (compare with the right panel of Fig.\ref{fig:sxem}).  The
right panels correspond to the empirical scaling law assuming $\Mg
\propto T^{1.94}$, consistent with the slope of the local \LxT
relation (Neumann \& Arnaud \cite{neumann01}).  The dispersion is
decreased even more as compared to the standard scaling.}
\label{fig:emsc}
\end{figure*}

\subsection{The \LxT relation and the empirical scaling law}

The bolometric cluster luminosity is given by:
\begin{eqnarray}
L_{\rm X} &= &\Lambda(T)\int_{0}^{\Rv} EM(r) 2\pi r
dr
\end{eqnarray}
where the cooling function, $\Lambda(T)$ varies as $\Lambda(T) \propto
T^{1/2}$.  For the standard self-similar model described above, the
bolometric luminosity follows the well known scaling relation:
\begin{eqnarray}
  L_{\rm X}& \propto & \Dz^{1/2} \ (1+z)^{3/2}\ T^{2}
  \label{eq:lxt}
\end{eqnarray}
which is inconsistent with the slope of the observed local \LxT
relation, $\alpha \sim 2.88$ (e.g Arnaud \& Evrard \cite{arnaud}).

As already mentioned in the introduction, we found evidence for a
steepening of the \MgT relation for hot clusters (Neumann \& Arnaud
\cite{neumann01}), in our previous study of the nearby cluster sample
considered here.  A gas mass varying as $\Mg \propto T^{1.94}$,
instead of $\Mg \propto T^{1.5}$, can both explain the observed \LxT
relation and significantly reduce the scatter in the scaled emission
measure profiles, when compared to the standard scaling.  In that case, the
emission measure scales with temperature as $EM \propto T^{1.38}$,
instead of $T^{0.5}$.  We will also consider this empirical scaling
law in the following section.  The dependence of the normalisation on
redshift and cosmological parameters remains a priori unchanged and we
will assume that the empirical slope of the relation does not evolve
with redshift, which is the simplest assumption.

\section{ Scaled emission measure profiles}

\subsection{Scaling procedure} 
As in
our previous studies, we scaled the emission measure profiles so that
they would lie on top of one another if obeying self-similarity.  The
scaled emission measure profiles, corresponding to the standard
scaling with $z$ and $T$ given Eq.~\ref{eq:em0sc}, are thus
defined\footnote{Note that the normalisation has been set so that the
central value would be unity for a $\betamodel$ with $\beta=2/3$ and
$\xc=0.123$, a $20\%$ gas mass fraction (see Eq.~\ref{eq:em0sc}) and
$\hc=1$.  All these factors are common to all profiles and their exact
value does not matter to check self-similarity} as:
\begin{eqnarray}
\widetilde{EM}(x)& =&
\Dz^{-3/2}(1+z)^{-9/2}\ \left(\frac{\kT}{10\
\keV}\right)^{-1/2}\nonumber \\
&& \times~\left(\frac{EM(r)}{4.10~10^{-6}{\rm cm^{-6}
Mpc}}\right) \nonumber\\
x& = &\frac{r}{\Rv}
\label{eq:em2emsc}
\end{eqnarray}
where $\Rv$ is defined by Eq.~\ref{eq:rv} and the emission measure is
derived from the surface brightness via Eq.~\ref{eq:sx2em}.

To introduce the empirical \EMT scaling relation ($EM \propto
T^{1.38}$), we simply have to introduce a corrective multiplicative
factor of $\propto T^{-0.88}$ to the previous equation:
\begin{eqnarray}
\widetilde{EM}(x)& =&
\Dz^{-3/2}(1+z)^{-9/2}\ \left(\frac{\kT}{10\
\keV}\right)^{-1.38}\nonumber \\
&& \times~\left(\frac{EM(r)}{6.0~10^{-6}{\rm cm^{-6}
Mpc}}\right) 
\label{eq:em2emscb}
\end{eqnarray}
For convenience, the corrective factor has been arbitrarily normalized
to 1 for a temperature equal to $6.5\ \keV$, which is the mean
temperature of the sample.

The scaled profiles depend on the assumed cosmological parameters, via
the angular distance $\dA$ used to convert angular radii to physical
radii and via the factor $\Dz = \Dzf$ appearing in the normalisation
of the profiles and of the \RvT relation.  The variation with redshift
of both quantities depends on $\Omo$ and $\Lambda$.  Therefore, if the
self-similar evolution model is valid, the scaled profiles of clusters
observed at various redshifts will coincide, but only for the correct
cosmological parameters.

The scaled profiles can thus be used both to check the validity of the
self-similar model and to put constraints on the cosmological
parameters, $\Omo$ and $\Lambda$.  This is described in detail in
Sect.~\ref{sec:test} and further discussed in Sect.~\ref{sec:disc}. 
To do so, we will use some general properties of the profiles that we
outline below.

\subsection{Scaled Profiles}

The scaled profiles are shown in Fig.~\ref{fig:emsc} for two
cosmological models, a critical density Universe $(\Omo=1)$ and a flat
model with $\Omo=0.3$ and $\Lambda=0.7$.  At first sight, distant
clusters appear remarkably similar to nearby clusters, once the
profiles are scaled.  As expected, the difference between the scaled
profiles, however, depends on the assumed cosmological parameters. 
The profiles plotted in the top and bottom panels of
Fig.~\ref{fig:emsc} are clearly different, in particular in the
relative position of the clusters for the different redshift ranges.

One further notes the large scatter in the scaled profiles in the
cluster core ($r < 0.1 \Rv$) and the remarkably common shape above
typically $0.1-0.2~\Rv$.  This was already noted for nearby clusters
(Neumann \& Arnaud \cite{neumann99}) and clearly also holds for
distant clusters.  The large scatter in the core is likely to be due
to cooling flows of various sizes.   For instance the four
clusters with the highest central value, MS 2137.3-2353, RXJ
1347.5-1145, MS 1512 and 3C295, are known, or suspected, to host
massive cooling flows.  The mass accretion rate, estimated using
standard cooling flow models, is as high as $3000 {\rm \msun/yrs}$ for
RXJ 1347.5-1145 (Schindler \etal~\cite{schindler97} and $500-900 {\rm
\msun/yrs}$ for 3C295 (Neumann~\cite{neumann99b}).  The presence of a
massive cooling flow in MS 1512.4+3647 is indicated by the detection
of luminous extended $H_{\alpha}$ emission (Donahue
\etal~\cite{donahue92}).  No detailed cooling flow analysis is
available for MS 2137.3-2353.  However, we note that the estimated
cooling time for this cluster is about 1/10 of the Hubble time (Allen
\& Fabian~\cite{allen98b}) and a clear drop of temperature is observed
in the center, similar to the one observed in RXJ 1347.5-1145 (Allen
\etal~\cite{allen01}).

A first quantitative check of similarity beyond the core can be made
by looking at the dispersion among the profiles at a given radius, for
the whole cluster sample.  The surface brightnesses are measured at
discrete values of the angular radii \footnote{Each data point is
actually the mean surface brightness in the radial bin considered and
not the surface brightness at the center of the bin as assumed here
and in the following.  We checked that the difference is negligible.}.
To compute the mean value and dispersion of the profiles at any
physical or scaled radius, a continuous profile was generated for each
cluster using a logarithmic interpolation of the data.

The scaling procedure always significantly reduces the differences
among the profiles, as can already be seen by comparing
Fig.~\ref{fig:sxem} and Fig.~\ref{fig:emsc}.  This is a first indication
that clusters obey scaling laws up to high redshift.  Let us for
instance consider the standard scaling with $\Omo=1$.  The relative bi-weight
dispersion of the emission measure profiles at a given radius is $\sim
100\%$ between 0.5 and 1 Mpc, whereas the dispersion drops to $\sim
40-45\%$ for the scaled profiles between $0.2$ and $0.5 \Rv$.

Furthermore, in the same range of radii, the scatter is further
decreased to $\sim 35-40\%$ for the empirical scaling relation ($EM
\propto T^{1.38}$, right panel).  This decrease is slightly more
pronounced in a low $\Omo$ Universe.  The improvement is not as
spectacular as for the local sample alone (a factor of 2 decrease of
the scatter).  However, this additional $T^{-0.88}$ scaling factor
introduces additional noise due to the uncertainties on the
temperatures.  These errors are particularly large for the distant
cluster sample.  The fact that there is still an improvement, in spite
of this additional noise, suggests that the empirical $\EMT$ scaling
relation fits better the cluster properties than the standard case, over the
redshift range $z=0.04-0.8$.  We will thus adopt this empirical
scaling relation in the following.

\section{Test of self-similarity and constraints on the cosmological
parameters}
\label{sec:test}

\subsection{Method}

Our aim is to check, in a quantitative way, the validity of the
self-similar model, and set constraints on the cosmological
parameters.  For that purpose, we need a better statistical estimator
than the calculated dispersion of the profiles at a given radius,
which we used in the previous section.  The relative dispersion is not
a global estimator and furthermore does not take into account
measurement errors.

We first derived, for each set of cosmological parameters, a scaled
reference profile, and an estimate of the intrinsic scatter around it,
using the nearby cluster sample data.  To do so, we estimated, at any
given scaled radius, the mean value of the different scaled $EM$
profiles, together with the corresponding standard deviation, at that
specific radius.  We computed this reference profile up to the radius
for which at least two nearby cluster profiles are still available. 
Note that measurement errors, which are much less than for the distant
cluster sample, can still contribute to the scatter.   Analytical
fits of the reference scaled profiles (for open and flat Universes) are
given in Appendix~\ref{sec:fit}.

We then considered the set of data points for the distant cluster
sample.  Each data point is the scaled emission measure
$\widetilde{EM_{i,j}}$ of cluster $j$, measured at the scaled radius
$x_{i,j}$, with corresponding $1~\sigma$ errors.  The error on the
temperature contributes to both the error on $x_{i,j}$ and on
$\widetilde{EM_{i,j}}$, while the error on the surface brightness
obviously only contributes to the later quantity.  These data points
are compared to the corresponding reference profile in the left panel
of Fig.~\ref{fig:zoom} for a critical density Universe.

If the self-similar model is valid, the distant cluster data points
after scaling must be consistent with the reference profile, within
the errors.  We thus computed the $\chi^{2}$ value of the distant
cluster data about the reference curve, for each cosmological model. 
This $\chi^{2}$ value can be used to assess in the standard way the
validity of the underlying self-similar model and to constrain the
cosmological parameters, considered as free parameters of the model.

The $\chi^{2}$ computation is not straightforward, because there are
non negligible errors on both variables $x$ and $\widetilde{EM}$,
these errors are correlated, and the reference curve is not linear. 
Furthermore, we have to take into account the existence of intrinsic
scatter.  The computation of $\chi^{2}$ is detailed in
Appendix~\ref{sec:chi2}.

\begin{figure}[t]
\epsfxsize= 8.5cm \epsfbox{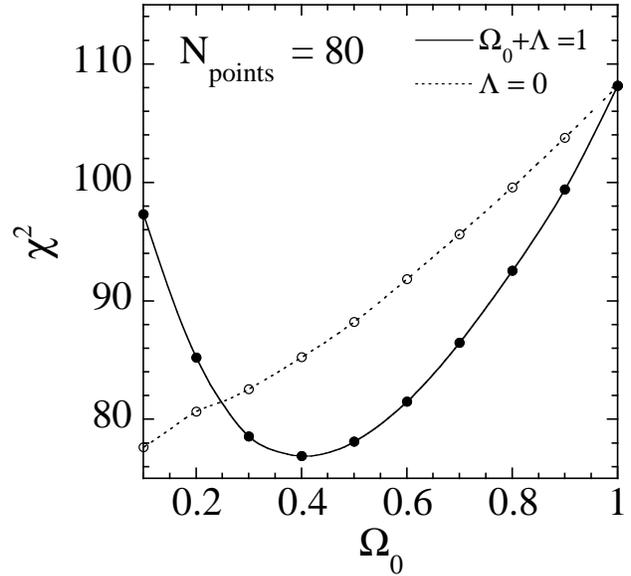} \caption{$\chi^{2}$ value of
the distant cluster data about the mean scaled emission measure
profile of nearby clusters, as a function of $\Omo$.  Full line: flat
Universe, $\Omo+\Lambda=1$.  Dotted line: Open Universe, $\Lambda=0$. 
}
\label{fig:chi2}
\end{figure}

Another technical issue is the choice of data points included in the
computation of the overall $\chi^{2}$.  First, and obviously, only
points for which there is a corresponding reference value from the
local sample can be included.  In practice, very few points are
excluded this way, since, for every cosmological model, the distant
clusters are usually traced up to smaller scaled radii when compared to
nearby clusters.  Furthermore, as discussed in the previous section,
the core properties are clearly dominated by different physics.  It is
thus better to exclude the central points to check self-similarity at
large radii and to constrain in a more significant way the
cosmological parameters.  For that purpose, including the central
points would be equivalent to add extra noise.  We thus considered a
fixed number of points, $N_{\rm p}$, defined as the $N_{\rm p}$ most
distant from the center in scaled coordinates.  Although the relative
position of the points depends somewhat on the cosmological
parameters, essentially the same data set is compared to the reference
curve in all cases\footnote{This would not be the case, if we had
considered a fixed region in terms of scaled radii.  The absolute
position of the profiles in the log-log plane is very sensitive on the
cosmology as can be seen by comparing the left and right panels of
Fig.~\ref{fig:zoom}.  A given angular radius corresponds to a smaller
scaled radius for a smaller value of $\Omo$.  This would have
introduced bias in the $\chi^{2}$ estimate, with more and more points
from the core included in the sample as $\Omo$ increases.}.  We both
considered $N_{\rm p}=80$ and $N_{\rm p}=150$ corresponding
respectively to a minimum scaled radius $x=0.2$ and $x=0.1$ for
$\Omo=1$.

\begin{figure*}[t]
\epsfysize=7.6cm \epsfbox{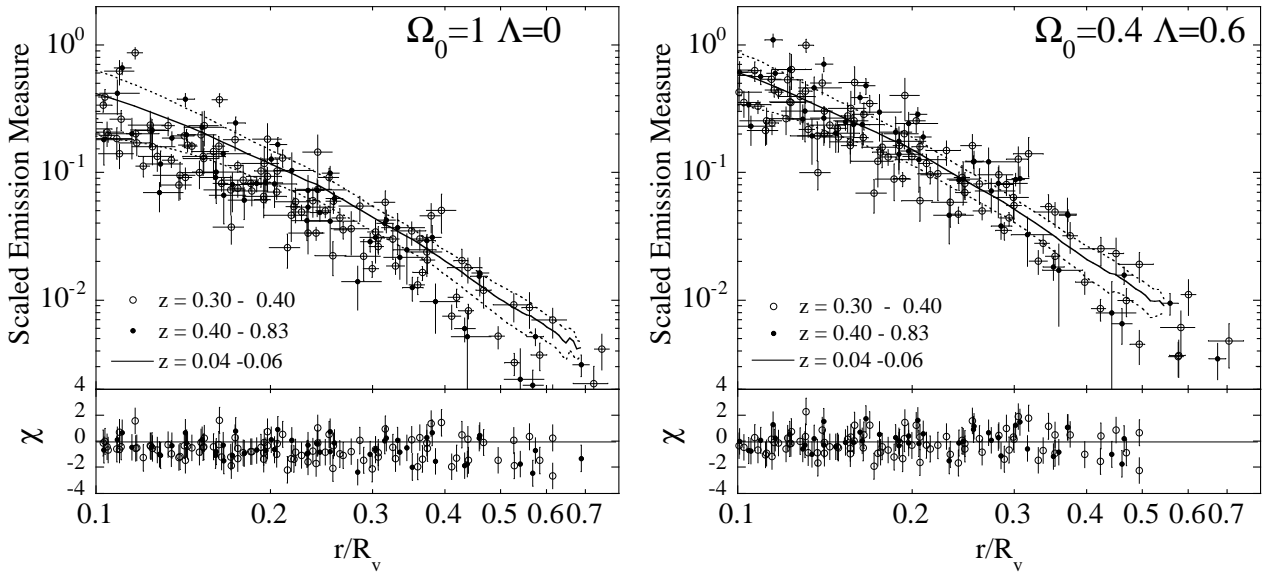} \caption{ Comparison between
the scaled emission measure profiles of the distant clusters (data
points) and the mean scaled profile determined from the nearby cluster
sample ($z=0.04-0.06$, full line) for two cosmological models.  The
profiles are scaled assuming $\Mg \propto T^{1.94}$.  The data for the
distant clusters are the same as shown in the right panels of
Fig.~\ref{fig:emsc} but each data point is now displayed individually
with $1\sigma$ error bars on both scaled variables.  Open and filled
points correspond to clusters in the redshift range $z=0.3-0.4$ and
$z=0.40-0.83$, respectively.  The dotted lines correspond to the mean
profile of nearby clusters, taken as a reference profile, plus or
minus the corresponding standard deviation.  The bottom panels show
the $\chi$ value of individual points about the reference profile,
taking into account the correlation between the errors and the
dispersion around the reference profile.  {\bf Left panel:} Results
for $\Omo=1, \Lambda=0$.  The distant cluster data are not consistent
with the nearby cluster data for a critical density Universe.  Most of
the points lie significantly below the reference profile.  {\bf Right
panel:} Results for $\Omo=0.4$ and $\Lambda=0.6$.  Note the good
agreement between the scaled emission measure profiles of distant
clusters and the mean scaled profile of nearby clusters.  There is
also no systematic variation of the $\chi$ value with radius, an
additional indication of self-similarity in shape.  }
\label{fig:zoom}
\end{figure*}

The variation of the $\chi^{2}$ value with $\Omo$ is plotted in
Fig.~\ref{fig:chi2} for a flat Universe $(\Omo+\Lambda=1)$ and an open
Universe ($\Lambda=0$) for $N_{\rm p}=80$.  The individual $\chi$
values for each data point are plotted on the bottom panels of
Fig.~\ref{fig:zoom} for a critical Universe (left panel) and for the
flat model, $\Omo=0.4$ and $\Lambda=0.6$, with the lowest $\chi^{2}$
value (right panel).

\subsection{Results}
For $\Omo=1$, most of the scaled distant cluster data points fall
significantly below the reference curve (Fig.~\ref{fig:zoom}, left
panel) with an overall $\chi^{2}$ value of 108 for $N_{\rm p}=80$
points.  Our data thus allow to exclude the critical density Universe
model at the $98\%$ confidence level.

Allowing $\Omo$ to vary, and considering first a flat Universe, we
find an excellent agreement of the distant cluster data with the
reference profile for a low density Universe (Fig.~\ref{fig:zoom},
right panel).  It indicates that hot clusters constitute a homologous
family up to high redshifts and strongly supports the underlying
self-similar model.  The smallest $\chi^{2}$ is reached for $\Omo=0.4$
($\Lambda=0.6$), with $\chi^{2}=77$ (reduced $\chi^{2}\sim 1$). 
Furthermore, no systematic variation of the $\chi$ value with radius
is observed, an additional indication of the self-similarity in shape
of the profiles (bottom right panel of Fig~\ref{fig:zoom}). 
Interestingly a strong constraint can be set on $\Omo$:
$\Omo=0.4^{+0.15}_{-0.12}$, with errors given at the $90\%$ confidence
level (corresponding to a $\Delta \chi^{2} = 2.7$).  This result is
not sensitive to the number of points considered.  For $N_{\rm
p}=150$, we obtain similar constraints $\Omo=0.43^{+0.13}_{-0.11}$,
with a slightly better reduced $\chi^{2}$ ($\chi^{2}=115)$.  The
constraint put on $\Omo$ for a flat Universe is remarkably consistent
with the constrain derived from luminosity distances to SNI:
$\Omo=0.3\pm 0.15$ at the $90\%$ confidence error only taking into
account statistical errors (Fig.  7 in Perlmutter
\etal~\cite{perlmutter}).

 We check the robustness of our results on the self-similarity of
clusters, with respect to the cluster temperature, by dividing the
distant cluster sample in two equal sub--samples.  We consider the
favored cosmological model ($\Omo=0.4,\Lambda=0.6$) and only data
points with $x > 0.08$ (corresponding to $N_{\rm p}=150$ data points
in total).  We obtain a reduced $\chi^{2}$ of 0.61 ($\chi^{2}=37$ for
$61$ data points) for the $\kT \leq 6.6~\keV$ subsample (13 clusters)
and a reduced $\chi^{2}$ of 0.88 ($\chi^{2}=78$ for $89$ data points)
for the $\kT > 6.6~\keV$ subsample (12 clusters).  Similarly,
splitting the sample with respect to the cluster redshifts, yields a
reduced $\chi^{2}$ of 0.82 ($\chi^{2}=80$ for $98$ data points) for
the $z~\leq~0.4$ subsample (15 clusters) and a reduced $\chi^{2}$ of
0.67 ($\chi^{2}=35$ for $52$ data points) for the $z>0.4$ subsample
(10 clusters).  In conclusion, for the favored cosmology, the distant
cluster data for each individual subsample are in excellent agreement
with the local reference profile.  This reinforces the validity of the
considered self-similar model, in particular the redshift dependence
of the scaling relations.

An open model ($\Lambda=0$) is also formally consistent with the data. 
However, the $\chi^{2}$ value keeps decreasing with decreasing $\Omo$,
preventing a strict definition of the constraints.  We thus only note
that for $\Omo=0.1$, we obtain a $\chi^{2}=78$, similar to the value
for the best model for the flat case.  All open models with $\Omo>
0.17$ give $\chi^{2}$ values which are larger than the values
corresponding to $90\%$ confidence range of the flat case. 
Furthermore, we cannot consider arbitrarily low $\Omo$ values. 
Obviously $\Omo$ must be greater than the baryonic density derived
from primordial nucleosynthesis ($\Omega_{\rm b} = 0.03-0.06$, Suzuki
\etal \cite{suzuki}).  In addition, the various approximations of the
scaling models (in particular to compute the over--density) become
less and less valid as $\Omo$ decreases.

\subsection{Origin of the constraint on the cosmological parameters}
\label{sec:origin}

Comparing the scaled emission measure profiles of clusters at
different redshifts appears to be a powerful method to constrain the
cosmological parameters.  To understand better the origin of the
constraint, we examine in more details the variation of the scaled
profiles, $\widetilde{EM}(x)$, with the cosmological parameters and
redshift.

It is useful to first explicitly identify this dependence, which is
somewhat complex.  The observed quantities are the surface brightness
profiles $S(\theta)$, which we correct for the $(1+z)^{4}$ dimming
factor.  Combining Eq.~\ref{eq:em2emsc} and Eq.~\ref{eq:sx2em},
together with Eq.~\ref{eq:rv} and identifying the relevant factors, we
can write:
\begin{eqnarray}
\widetilde{EM}(x)& \propto & \Dz^{-3/2}\ \left(1+z\right)^{-9/2}
\left[S(\theta)\left(1+z\right)^4\right] \\
x &  \propto &  \theta\ \dA  \Dz^{1/2}\  \left(1+z\right)^{3/2}
\label{eq:cosmo}
\end{eqnarray}
Scaling the observed and dimming corrected $S(\theta)$ profile
corresponds to translating it in a log-log plane.  On the one hand,
there is the translation of $\log(\dA)$ along the x direction related
to the conversion of angular radius into physical radius.  On the other hand,
the cluster cosmological evolution requires an additional translation
of $ -(3/2)\ \log(\Dz) - (9/2)\ \log(1+z)$ in the y direction, and of
$ (1/2)\ \log(\Dz) + (3/2)\ \log(1+z)$ in the x direction, i.e along a
line of slope -3.

The cosmological parameters appear in the angular distance $\dA$ and
through the cluster over--density factor $\Dz$.  In the log-log plane,
the scaled profiles for two different cosmological models simply
differ by translations.  At a given redshift, varying the cosmological
parameters simply corresponds to the same translation in the log-log
plane for all the scaled profiles.  The cosmological parameters can
thus only be constrained by comparing profiles at different redshifts.

\begin{figure}[t]
\epsfxsize=8.5cm \epsfbox{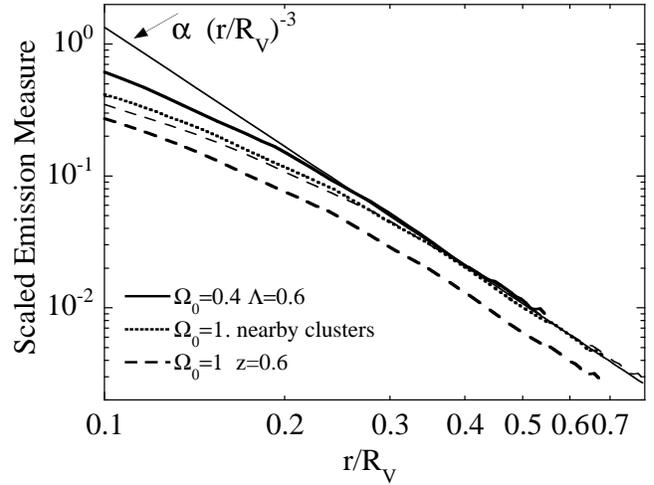}
\caption{Dependence of the scaled profiles on the cosmological
parameters assumed.  Thick full line: reference scaled profile from the
nearby cluster sample derived for $\Omo=1$.  Dotted line: same for
$\Omo=0.4$, $\Lambda=0.6$.  At large radii, the logarithmic slope of
the profile is $\sim -3$ (thin full line), corresponding to a \betamodel
with $\beta=2/3$.  Thick dashed line: Scaled profile of a $z=0.6$ cluster
one would derived assuming $\Omo=1$, if the `true' cosmological model
was $\Omo=0.4$, $\Lambda=0.6$, i.e if the `true' scaled profile would
follow the dotted line.  Thin dashed line: same only taken into account
intrinsic cosmological evolution.  By using wrong values of the
cosmological parameters, the derived profiles of distant and nearby
clusters do not coincide any more.  The discrepancy is essentially due
to the dependence of the derived profiles on the angular distance
assumed (see text for full discussion).  }
\label{fig:refo}
\end{figure}

The reference scaled profile is determined from nearby cluster data. 
This profile depends itself on the cosmological parameters.  At low
redshifts, increasing $\Omo$ (for both flat and open models) mainly
affects the $\Dz$ factor, the angular distance being almost
insensitive to the cosmological parameters.  As $\Dz$ increases with
$\Omo$, increasing $\Omo$ moves the scaled profile down and to the
right, along the line of slope -3 (defining the scaling translation
due to cosmological cluster evolution).  This is illustrated in
Fig.~\ref{fig:refo}, where we compare the reference profile from the
nearby cluster sample for two flat cosmological models ($\Omo=0.4$,
full line, and $\Omo=1$, dotted line).  A remarkable feature is the
coincidence of the scaled profiles for the two models at large radii. 
This is due to the coincidence between the slope of the scaling
translation (-3) and the slope of the profile at large radii (thin
line in Fig.~\ref{fig:refo}), so that the profile is just translated
`along it self'.  Note that this slope at large radii simply
corresponds to a $\betamodel$ with $\beta=2/3$, which was shown to fit
well the mean profile of nearby clusters (Neumann \& Arnaud
\cite{neumann99}).  At smaller radii, the slope of the profile becomes
smaller than -3.  As a result, the scaled profile for a high $\Omo$
value always lies below the corresponding profile for a lower value of
$\Omo$ (Fig.~\ref{fig:refo}).

Let us now consider a high redshift cluster and assume that the
correct cosmological model is a flat Universe with $\Omo=0.4$.  The
scaled profile of this cluster, derived for $\Omo=0.4$, will follow
the corresponding reference curve.  However, this will not be the case
if we assume another $\Omo$ value.  $\Dz$ varies more rapidly with
$\Omo$ as the redshift increases.  As a result, the scaled profile of
a high redshift cluster is more affected by a change of $\Omo$ than
the scaled profile of a lower redshift cluster.  Taking {\it only}
into account the translation related to cosmological evolution, we
compare the scaled profile of a $z=0.6$ cluster, one would obtain
assuming $\Omo=1$ (thin dashed line in Fig~\ref{fig:refo}) to the
corresponding reference curve obtained for this cosmological
parameter.  The profiles still coincide at large radii, for the reason
explained above.  At small radii the $z=0.6$ profile is below the
reference curve.  It must be noted that this effect is small above 0.1
virial radius (less than $20\%$), i.e in the radial range considered. 
However, at high redshifts, we have also to take into account the
variation of the angular distance with $\Omo$, which decreases with
increasing $\Omo$.  The profile has to be further moved to the left,
along the x axis, by
$\log(\dA_{(\Omo=1)}/\dA_{(\Omo=0.4)})$\footnote{It is thus still
below the profile corresponding to a lower $\Omo$ value.  It can be
also shown that in Eq.~\ref{eq:cosmo} the product $\Dz^{1/2}\ \dA$
(proportional to the inverse of the angular virial radius) always
increases with increasing $\Omo$, and the profile remain globally
moved to the right.  The direction of the translation, down and right,
is readily apparent when comparing individual data points in the left
and right top panels of Fig.~\ref{fig:zoom}.}.  At all radii, the
profile of the $z=0.6$ cluster (thick dashed line in
Fig~\ref{fig:refo}) does not coincide anymore with the reference
profile.  For a profile shape varying roughly as $x^{-3}$ at large
radii, the effect of $\dA$ on the scaled emission measure at a given
scaled radius is large, $\propto \dA^{3}$.  At $z=0.6$, the angular
distance is about $18\%$ higher for $\Omo=1$ than for $\Omo=0.4$ and
the profile is $\sim 60\%$ below the corresponding reference profile.

This is exactly what we observe in Fig.~\ref{fig:zoom}.  For
$\Omo=0.4$ (right panel) distant cluster profiles are consistent with
the reference curve.  For $\Omo=1$ (left panel), all the data are
moved down.  The decrease is more important for distant clusters,
which are now systematically lower than the reference curve of nearby
clusters, allowing us to exclude this cosmological model.  The same
reasoning applies for an open Universe.  However, the variation of
$\Dz$ and $\dA$ with $\Omo$ is less pronounced for an open Universe
than for a flat Universe.  The differential effect is less important
and the profiles coincide only for lower values of $\Omo$.

In conclusion, increasing $\Omo$ moves the scaled profile of a given
cluster down and right and decreases the scaled emission measure at a
given scaled radius.  However, in the radial range considered, the
derived scaled profiles of any distant cluster, {\it when compared} to
the corresponding reference profile, is mostly sensitive to the
angular distance $\dA$ at the cluster redshift.  This is this
dependence, which essentially allow to constrain the cosmological
parameters, via the well known dependence of $\dA$ with $\Omo$ and
$\Lambda$.

\section{Evolution of the Lx-T relation}

\begin{figure*}[t]
\epsfxsize=17.5cm \epsfbox{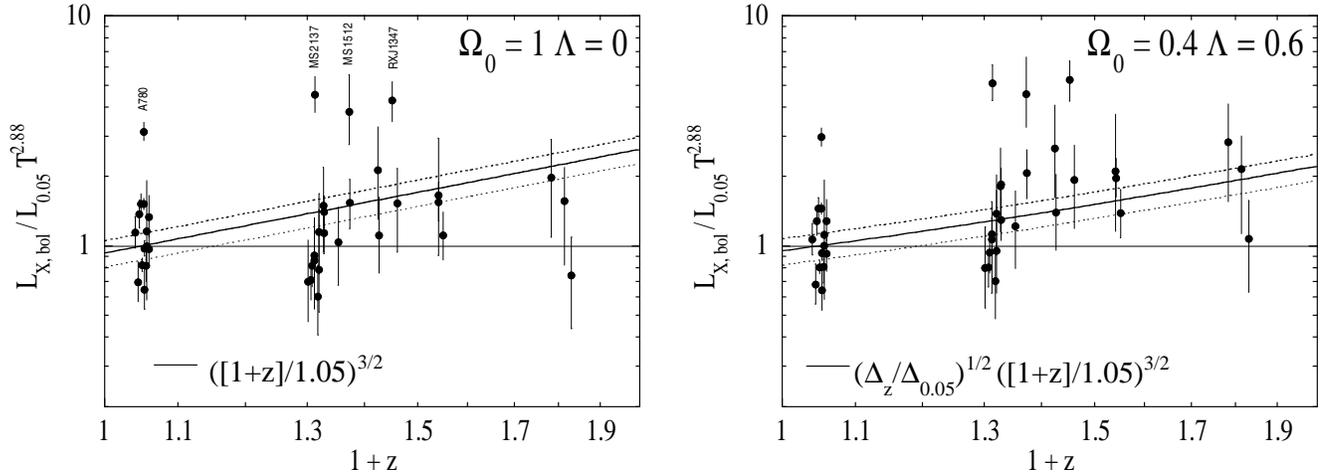} \caption{ Evolution of the
normalisation of the \LxT relation for a critical universe (left
panel) and a flat Universe with $\Omo=0.4$ (rtight panel).  Data
points: Observed bolometric luminosity divided by the luminosity
estimated from the local \LxT relation and the cluster temperature. 
Error bars include errors on the luminosity and temperature.  Full
line: theoretical evolution for the chosen cosmology and the
self-similar model of cluster formation.  Dotted line: the theoretical
curve offset by the intrinsic scatter in $\log(L_{\rm X})$ estimated
by Arnaud \& Evrard (\cite{arnaud}).  The evolution of the
normalisation of the \LxT relation is consistent with the self-similar
model for ($\Omo=0.4$,$\Lambda=0.6$).}
\label{fig:LxT}
\end{figure*}

We found that the scaled $EM$ profiles of distant clusters coincide
with the profile of nearby clusters, using a flat cosmological model
with $\Omo=0.4$.  This means that the surface brightness profiles of
distant clusters follow the evolution with redshift expected in the
self-similar model, for this set of parameters.  Since the X--ray
luminosity is nothing else than the integral of the surface brightness
profile, the evolution of the \LxT relation should therefore also
comply with this model.  We check this point now.

For consistency, we use the bolometric cluster luminosity estimated
from the ROSAT data presented here, rather than ASCA. For each
cosmological model, the total emission measure within the virial
radius is estimated by integrating the profiles up to the detection
radius.  The contribution beyond that radius was estimated using a
\betamodel with a slope $\beta=2/3$, normalized to the emission
measure at 0.3 virial radius.  The luminosity was then estimated using
the cooling function computed with a MEKAL model at the cluster
temperature (Tab.~\ref{tab:data}).  The ROSAT luminosity values,
computed for $\Omo=1$ are given in Tab.~\ref{tab:data}.  They are in
good agreement with the corresponding ASCA estimates from the
literature.  The median ratio between the two estimates is 0.97, with
a standard deviation of $0.2$ and there is no specific trend with
redshift.

The considered self-similar model assumes a non-evolving slope of the
\EMT scaling relation, consistent with the slope (2.88) of the local
\LxT relation established by Arnaud \& Evrard (\cite{arnaud}).  We
thus study the evolution of the normalisation of the \LxT relation,
assuming a constant slope of 2.88.  For each cluster, we define, as in
Sadat \etal~(\cite{sadat}), the quantity:
\begin{eqnarray}
	C_{\rm obs} &=&\frac{L_{\rm X}}{L_{0.05}\ T^{2.88}}
\label{eq:cz}
\end{eqnarray}
where $L_{\rm X}$ is the measured bolometric luminosity and $L_{0.05}$
is the normalisation of the local \LxT relation, taken from the nearby
sample (excluding A780, see below).  This normalisation is perfectly
consistent with the data of Arnaud \& Evrard (\cite{arnaud}), the
ratio of the two normalizations is 0.99 (for $\Omo=1$).  From
Eq.~\ref{eq:lxt}, this quantity should evolve as:
\begin{eqnarray}
	C_{\rm mod}(z) = \left(\frac{\Dz}{\Delta_{0.05}}\right)^{1/2}\
	\left(\frac{1+z}{1.05}\right)^{3/2}
\end{eqnarray}
For consistency, we have normalized the theoretical function, to the
value at $z=0.05$, the median redshift of the nearby sample.  The
observed $C_{\rm obs}$ values, with error bars estimated from the
errors on luminosity and temperature, are compared to the theoretical
curve, $C_{\rm mod}(z)$, in Fig.~\ref{fig:LxT} for a critical density
Universe (left panel) and for our best fit model ($\Omo=0.4,
\Lambda=0.6$, right panel).  Four clusters (A780, MS2137, MS1512 and
RXJ1347) stand out with particularly high luminosities.  Strong
cooling flow clusters are known to lie above the \LxT relation of weak
or non-cooling flow clusters, considered by Arnaud \& Evrard
(\cite{arnaud}).   As mentioned above, MS2137, MS1512 and RXJ1347
are indeed known to host strong cooling flows and the same is true for
A780.  They are thus discarded in the following.

For a critical density Universe most of the data points lie below the
theoretical curve, indicating that the distant clusters are
under-luminous as compared to the theoretical expectation.  This is
the same effect as observed for the scaled emission measure, which was
also found to be too low as compared to the reference curve.  We
computed the $\chi^{2}$ value of the distant cluster data points about
both the theoretical curve and a constant value of 1, corresponding to
no evolution.  We assume an intrinsic scatter of 0.13 in $\log(L_{\rm
X})$ (Arnaud \& Evrard \cite{arnaud}).  We obtained a reduced
$\chi^{2}$ of 2.3 and 1.1, respectively.  We thus note that the
observed $C_{\rm obs}$ values seem actually consistent with no evolution at
all, as found in previous studies of the \LxT relation for $\Omo=1$
(Sadat \etal \cite{sadat}; Fairley \etal \cite{fairley}; Reichart
\etal \cite{reichart}).

On the other hand, the distant cluster data are consistent with the
expected evolution of the \LxT relation, for $\Omo=0.4, \Lambda=0.6$. 
The $\chi^{2}$, in this case, is $\chi^{2}=25$ for 22 clusters
(reduced $\chi^{2}_{\rm red}=1.1$).  This good $\chi^{2}$ also
indicates that the data are consistent with no evolution of the slope
of the \LxT relation with $z$, as we assume.  By comparison, a reduced
$\chi^{2}$ of $\chi^{2}_{\rm red}=2.0$ when the data are compared with
the no evolution curve.  The origin of the improvement, as compared to
the $\Omo=1$ model, is a combination of two factors i) the measured
luminosity is higher for lower $\Omo$, because the estimated distance
is larger and ii) the expected evolution of the \LxT relation is more
modest due the factor $\Dz$, a decreasing function of $z$.  We find that $C_{\rm mod}(z)$, for this
cosmological model, is well approximated by a power law over the
redshift range considered here ($z=0.-0.83$): $C_{\rm mod}(z) \cong
(1+z)^\eta$, with $\eta=1.23$.  This can be compared with the results
of Reichart \etal (\cite{reichart}), who assume such a dependence. 
Their results are based on a compilation of data from the literature
for clusters with $z<0.5$.  For $\Omo=0.4, \Lambda=0.6$ ($q_{o}=-0.4$)
they found $\eta = 1.36^{+0.54}_{-1.22}$, their best fit is thus also
in good agreement with the theoretical expectation.

In summary, as expected from our study of the emission measure
profiles, the normalisation of the \LxT relation, derived for
a flat low density Universe, does evolve with redshift.  The observed
evolution is consistent with the self-similar model of cluster
formation.

\section{Discussion}
\label{sec:disc}

\subsection{Cluster formation and evolution}

We found an excellent agreement, both in shape and normalisation, of
the set of scaled profiles of distant clusters with the reference
nearby scaled, for a low density flat Universe.  This indicates that
hot galaxy clusters constitute a homologous family up to high
redshift, with the cluster properties scaling with $z$ as expected in
the simplest self-similar model.  These scaling laws are derived from
the assumption that clusters form at fixed density contrast as
compared to the critical density of the Universe.  Our results thus
support this standard picture for the gravitational collapse of the
dark matter component.

Consistently, the evolution of the normalisation of the \LxT relation
was found to comply with the self-similar model for $\Omo=0.4$,
$\Lambda=0.6$.  The apparent inconsistency with this model, claimed in
some previous studies, was in fact due to the a priori choice of a
particular cosmological model ($\Omo=1$), and is not, per se, an
indication of extra-physics.  That does not mean that such physics
does not exist.  We emphasize again that the present study only
concerns relatively hot clusters, for which non-gravitational effects
like pre-heating is minimal, and that the slopes of the \MgT and \LxT
relations remain inconsistent with the standard scaling laws.  The
simplest self-similar model is clearly insufficient to fully describe
the properties of the gas component of the ICM. Our data are
consistent with no evolution of these slopes with $z$ but much better
temperature estimates and larger samples are required for further
tests of this point.  If this was confirmed, the empirical slopes,
together with similarity in shape up to high $z$, have to be explained
in terms of the specific gas physics in the picture of structure
formation.

\subsection{A new method to constrain the cosmological parameters from
clusters?}

The basic assumption of the method to constrain the cosmological
parameters, that we validate a-posteriori, is that clusters form a
homologous population.  In particular, once scaled according to
cluster temperature and redshift, the $EM$ profiles, derived from the
observed surface brightness profiles $S(\theta)$, follow a universal
function, common to all clusters.  This universal function, determined
from nearby cluster data, is then used as `standard candle' to
constrain the cosmological parameters.  The scaled profiles are not
used as simple distance indicators, in the strict sense.  They do
depend on the angular distance, via the conversion of angular radius
into physical radius, but, in the self-similar model, the evolution of
the scaling relations used in the scaling process also depends on the
cosmological parameters.  However, we have shown that the second
effect is actually negligible.  The scaled profiles show a strong
dependence on the angular distance, $\propto \dA^{3}$ at any given
scaled radius.  This makes them competitive with other distance
indicators like SNI ($\propto \dA^{2}$) and indeed, we were able to
derive constraints of similar quality.

The method is however by no mean as powerful as the cosmological test
based on the evolution of the cluster mass function, $N(M,z)$, which
depends much more strongly on $\Omo$.  For instance, the abundance of
massive clusters falls by an order of magnitude by $z\sim 0.6$ in a
critical density Universe, while it remains unchanged in an $\Omo=0.3$
Universe (Blanchard \& Bartlett~\cite{blanchard}).  However, indirect
mass indicators, like $T$ or $L_{\rm X}$, have to be used and the
selection function for flux-limited cluster surveys depends on the
cluster scaling and structural properties.  This $N(M,z)$ test, by
nature, thus requires a good knowledge of the cluster scaling
properties, and their evolution with $z$, that we study here.  The
major advantage of the present method is that it is more direct, i.e
directly based on observed quantities.

The proposed method suffers, nevertheless, from intrinsic limitations. 
First, there is some intrinsic dispersion in the cluster properties. 
The typical dispersion is about $\pm 25\%$, which is not so different
from the $\sim 60\%$ in $\dA^{3}$ between an $\Omo=0.4$
($\Lambda=0.6$) and an $\Omo=1$ cosmology for a cluster at a redshift
as high as $z=0.6$.  Therefore, the method requires to consider a
large sample of clusters, the scaled profiles of distant clusters will
coincide with the reference profile, for the correct values of the
cosmological parameters, only in a statistical sense (i.e. on average)
and not on a cluster by cluster basis.  However, this intrinsic
dispersion can be measured and explicitly taken into account.  This is
done in the present analysis, where the set of distant cluster
profiles is compared to the reference profile, with its dispersion,
using $\chi^{2}$ statistics.

A more serious concern is that the method is intrinsically model
dependent.  As for other distance indicators, we cannot exclude that
some evolutionary effects, not taken into account, bias the results. 
In particular, the present analysis relies on the assumption that the
slope of both the \EMT and \RvT relations does not evolve with $z$. 
If this was not the case, different constrains on $\Omo$ might be
obtained.  For instance, a decrease of the \EMT relation slope with
$z$ would boost the scaled emission measure profiles of high $z$
clusters, as compared to low $z$, and could possibly mimic the effect
of a low $\Omo$ Universe.   To further quantify this point, we let
the slope of the \EMT relation vary with redshift.  As a test case, we
consider a very simple model.  With respect to the standard \EMT
relation, $EM\propto \Dz^{3/2}(1+z)^{9/2} T^{1.38}$, the \EMT
relations at the various redshifts are rotated, around a common
reference point of given temperature $T_{\rm p}$, so that the slope
varies linearly with $z$: $EM' = EM (T/T_{\rm
p})^{-(\delta\alpha(z))}$.  The slope at $z=0.05$, the median redshift
of the nearby cluster sample is kept unchanged and we consider a slope
change as high as $\pm 20\%$ at $z=0.8$: $\delta\alpha(z) = \pm
0.37(z-0.05)$, corresponding to $EM \propto T^{1.1}$ or $EM \propto
T^{1.66}$ at $z=0.8$.  The scaled emission measure profiles are then
given by Eq.~\ref{eq:em2emscb}, modified by a multiplicative factor of
$(T/T_{\rm p})^{-\delta\alpha(z)}$.  The change in the \EMT relation,
and thus its impact on the $\Omo$ estimate, depends somewhat
critically on $T_{\rm p}$; it is obviously maximal for low values of
$T_{\rm p}$.  If we take as reference temperature, $T_{\rm
p}=3.5~\keV$, the minimum temperature of the sample, the scaled
emission measure of a $10~\keV$ cluster at $z=0.8$ is increased by
about $ 35\%$, for a $20 \%$ decrease in slope.  As expected, higher
$\Omo$ values are then derived: $\Omo=0.6^{+0.17}_{-0.15}$ (flat
Universe).  If instead we assume a steepening of the \EMT relation
with $z$, we obtain $\Omo=0.25\pm0.11$.  Both evolution models are
consistent with the data, the corresponding reduced $\chi^{2}$ is not
significantly different from the one obtained for the standard model. 
We note however that, in both cases, a critical density Universe
remains excluded.  Furthermore, when we take $T_{\rm p}= 6.5~\keV$,
the median temperature of the sample, the slope evolution has a
negligible effect on the derived parameters: we obtain
$\Omo=0.47\pm0.14$ and $\Omo=0.34^{+0.15}_{-0.11}$ for a $+/- 20\%$
change in slope, respectively.  This systematic effect is twice
smaller than the statistical errors on the $\Omo$ value.  Finally, we
also checked the effect of a similar change in the slope of the \RvT
relation.  The effect is similarly negligible: we obtain
$\Omo=0.30^{+0.15}_{-0.09}$ and $\Omo=0.47^{+0.15}_{-0.13}$ for a
$20\%$ increase and decrease of the slope respectively.

Our conclusions also depend on the assumed \EMT relation, because we
are not considering clusters of similar temperatures at all redshifts. 
The median temperature is $5.8~\keV$ for the nearby sample, and
$6.6~\keV$ and $7.9~\keV$ for distant clusters in the redshift ranges
$0.3$--$0.43$ and $0.45$--$0.83$, respectively.  For a shallower \EMT
relation, like the standard scaling relation, most of the scaled
profiles of distant clusters, lie above the scaled profiles of nearby
cluster for $\Omo=0.3$ (Fig.~\ref{fig:emsc}).  In that case, a better
agreement between the distant cluster data and the nearby cluster data
is obtained for $\Omo=1$ than for $\Omo=0.3$.  However, we emphasize
again that the standard scaling law is not consistent with the slope
of the \LxT relation and that the adopted scaling relation decreases
significantly the scatter of the scaled profiles.

Nevertheless, a precise determination of the scaling with temperature,
and of its possible evolution, remains essential to achieve a fully
consistent description of cluster evolution and to assess possible
systematic errors on the cosmological parameters.  Again, the quality
of the present data is rather insufficient for high precision tests. 
Finally, there will always remain the possibility of a degeneracy between
the evolution of the cluster properties and the variation of the
angular distance with redshift.  Such systematic errors can only be
assessed by comparing the results obtained by various methods.  The good
agreement obtained between our results and the constraints based on
the luminosity distance of SNI is an encouraging sign that both
methods are unbiased and the underlying models correct.

\subsection{Comparison with previous work based on the
size-temperature relation}

The cosmological parameters can also be constrained, as proposed by
Mohr \etal (\cite{mohr00}), using the isophotal size - temperature (ST)
relation.  Mohr \etal (\cite{mohr00}) showed that the normalisation of
this relation is insensitive to cluster cosmological evolution,
considering the same model than in the present study.  Their test of
the cosmological parameters is thus made via the dependence of the size
on the angular distance.  Both this method and ours thus use
quantities derived from cluster surface brightness profiles, as
distance indicators. 

The main difference between the two methods is that we consider scaled
quantities rather than physical quantities.  At large radii,
considered by Mohr \etal (\cite{mohr00}), this is equivalent.  Due to
the coincidence between the slope of the scaling translation related
to cluster cosmological evolution and the slope of the profiles (see
Sect.~\ref{sec:origin}), the profiles of all clusters coincide at
large radii, both in the scaled space and in the physical space.  This
is the origin of the invariance of the isophotal size with redshift
(the arguments developed by Mohr \etal~(\cite{mohr00}) are actually
similar).  For an isophotal size evaluated from cluster images, the
method of Mohr \etal~(\cite{mohr00}) is equivalent, in our approach,
to consider only data points at a given scaled emission measure.

Our method, where we consider the whole set of data points, can thus
be regarded as a generalization of the method proposed by these
authors.  Note that, by working with scaled quantities, we are able to
consider data at small radii, where cosmological evolution has to be
taken into account (even if it does not depend sensitively on the
cosmological parameters).  The method we propose presents several
advantages.  Obviously tighter constrains can be obtained by
considering the whole set of data points.  No parametric fit of the
surface brightness profiles, as the one introduced by Mohr \etal
(\cite{mohr00}), is required.  Furthermore, it allows a more complete
test of the underlying self-similar model.

We stress on the agreement between the results obtained Mohr \etal
(\cite{mohr00}) and ours.  Both studies exclude a critical density
Universe.  Low $\Omo$ values are favored, somewhat lower when we use
the evolution of the ST relation.  One also notes that lower $\Omo$
values are favored for an open model than for a flat model.

\section{Conclusion}

In this work based on ROSAT data and published ASCA temperatures we
study the surface brightness profiles of a sample of hot ($\kT >
3.5~\keV$) galaxy clusters, covering a redshift range $z=0.04-0.83$. 
For both open and flat cosmological models, the derived emission
measure profiles are scaled according to the self-similar model of
cluster formation.  We use the standard scaling relations of cluster
properties with redshift.  The physical radius is normalized to the
virial radius, estimated from the classical virial relation.  The
slope of the \EMT relation depends on the assumed slope of the \MgT
relation.  We consider both the standard scaling relation $\Mg \propto
T^{1.5}$ and the empirical local relation $\Mg \propto T^{1.94}$
(Neumann \& Arnaud (\cite{neumann01}), assuming the slope does not
evolve with $z$.

Our analysis of the scatter of the scaled profiles, suggests that the
empirical slope of the \MgT relation fits better the cluster properties
than the standard value, over the whole redshift range $z=0.04-0.8$.  As for
nearby clusters, a large dispersion in the central core is observed,
and we therefore consider only the region above typically $0.1 \Rv$.

Applying the empirical \EMT relation, the set of scaled profiles of
the distant cluster sample are compared to the average scaled profile
of nearby clusters, using a $\chi^{2}$ test.  An excellent agreement,
both in shape and normalisation, of the distant cluster data with this
reference nearby scaled profile is obtained for a flat low density 
Universe (see also below).  Consistently, the evolution of the
normalisation of the \LxT relation was found to comply with the
self-similar model.  The apparent inconsistency with this model,
claimed in some previous studies, was in fact due to the a priori
choice of a particular cosmological model ($\Omo=1$).

This indicates that hot galaxy clusters constitute a homologous family
up to high redshifts and supports the standard picture for the
gravitational collapse of the dark matter component.  However, the
simplest self-similar model is insufficient to fully describe the
properties of the gas component of the ICM, as indicated by the
non-standard slope of the \MgT (and \LxT) relation.  If confirmed,
this slope, together with similarity in shape up to high $z$, have to
be explained in terms of the specific gas physics in structure
formation scenario.

Because of the intrinsic regularity of the hot cluster population, we
showed that the scaled emission measure profile, determined from
nearby cluster data, can be used as `standard candle' to constrain the
cosmological parameters, the correct cosmology being the one for which
the profiles at different redshifts coincide.  The scaled profiles of
distant clusters, as compared to the reference profile, mostly depend
on the angular distance, as $\dA^{3}$, making them powerful distance
indicators.  The method is, in addition, more powerful than the test
based on the size--temperature relation (Mohr \etal~\cite{mohr00}),
because it utilizes the full information contained in the cluster
profiles, rather than a particular point of the profiles.

Using this new method, we were able to exclude a critical-density
model ($\Omega_0=1$) (at 98\% confidence level).  The data favor a
flat Universe with a low matter density, even if the open model is not
formally excluded.  We find a value of $\Omega_0=0.40^{+0.15}_{-0.12}$
(at 90\% confidence level).  This test relies on the fact that we are
using the right scaling relations, in particular for the \EMT
relation.  It is thus, by nature, a model dependent method, although
the model can be to some extent, validated a posteriori.

At this stage, our proposed method has to be taken more in terms of an
independent consistency check of the constraints on cosmological
parameters rather than ``an ultimate cosmological test''.  The
constraint derived on $\Omo$ is in remarkable agreement with the
constraint obtained from luminosity distances to SNI (Perlmutter \etal
\cite{perlmutter}) or from combined analysis of the power spectrum of
the 2dF galaxy redshift Survey and the CMB anisotropy (Efstathiou
\etal~\cite{efstathiou};  see also
Melchiorri \etal~\cite{melchiorri}; Stompor \etal~\cite{stompor}; de
Bernardis \etal~\cite{debernardis}; Pryke \etal~\cite{pryke} and
references therein).  This is an additional sign that we are entering
an era where cosmological tests converge and we can expect that soon
the cosmological parameters will be accurately known.  In this
context, cluster scaling and structural properties will be more
adapted to test the physical processes in the structure formation
picture.  Significant progresses in this field require high quality
data with measurements down to the virial radius that will be provided
by the new generation of X-ray observatories (Chandra and XMM-Newton). 
They also require a large sample of distant and nearby clusters so
that i) the intrinsic dispersion is pinned down, ii) we improve our
knowledge of the local relations and the temperature and dark matter
profiles, and iii) we fully assess the evolution with $z$.

\acknowledgements We thank A.Blanchard for his participation to the
early stage of the study and J.Ballet for very useful discussions on
the statistical analysis of the data.  We thank the referee for useful
suggestions.  This research has made extensive use of the NASA's
Astrophysics Data System Abstract Service; the SIMBAD database
operated at CDS, Strasbourg, France; the NASA/IPAC Extragalactic
database (NED); the High Energy Astrophysics Science Archive Research
Center Online Service, provided by the NASA/Goddard Space Flight
Center and the MPE ROSAT Public Data Archive.

\appendix

\section{Analytical fit of the reference scaled profiles}
\label{sec:fit}

 For a critical Universe the reference profile derived from the set of
nearby cluster profiles is well fitted by a \betamodel above a scaled radius
of $x=0.1$:

\begin{eqnarray}
\widetilde{EM}(x)& =&\widetilde{EM}_{0} \left[1+\left(\frac{x}{x_{\rm
c}}\right)^{2}\right]^{-3\beta+0.5}
\label{eq:fit}
\end{eqnarray}
with $\widetilde{EM}_{0}=0.945$, $x_{\rm c}=0.116$ and $\beta=0.67$. 
This analytical formula is accurate to $\sim 7\% $ (with typical
errors less than $5\%$) in the range $x=0.1$--$0.7$.  We emphasize that
this formula {\it must not be used} for lower values of x (where the
scatter increases and the profiles are on average more peaked), as well
as above $x=0.7$, corresponding to the last measured point.

The individual cluster scaled profiles, derived from the observed
surface brightness profiles, depend on the cosmological parameters, via
the factor $\Dz$ and the angular distance, as given in
Eq.~\ref{eq:cosmo}.  The reference profile, for any cosmological
model, is well approximated by simply scaling the $\Omo=1$ reference
profile, with the $\Dz$ and $\dA$ factors estimated at $z=0.05$, the
mean redshift of the sample.  The reference profile is thus given by a
\betamodel (Eq.~\ref{eq:fit}) with:
\begin{eqnarray}
   \widetilde{EM}_{0}&=&0.945~\Delta_{0.05}^{-3/2}\\
   x_{\rm c}&=&0.116~\Delta_{0.05}^{1/2}~f_{\rm {d_{A}}}\\
   \beta&=&0.67
\end{eqnarray}
where $f_{\rm {d_{A}}}$ is the angular distance normalised to its value
for $\Omo=1$. The $\Delta_{0.05}$ factor is given Eq.~\ref{eq:dz} and
Eq.~\ref{eq:dc}, with $z=0.05$.  It is accurately (within less than
$0.2\%$) aproximated by the following polynomial expression:
\begin{eqnarray}
    \Delta_{0.05} &=&1 - 0.365(\Omo-1)-0.174(\Omo-1)^{2}~{\rm
    for}~\Omo<1,\Lambda=0 \nonumber \\
	\Delta_{0.05} &=&1 - 0.522(\Omo-1)-0.199(\Omo-1)^{2}~{\rm
	for}~\Omo+\Lambda=1 \nonumber \\
\end{eqnarray}
Similarly the $f_{\rm {d_{A}}}$ factor can be aproximated by:
\begin{eqnarray}
    f_{\rm {d_{A}}} &=&1 + 1.25~10^{-2}(\Omo-1)~{\rm
    for}~\Omo<1,\Lambda=0 \nonumber \\
       f_{\rm {d_{A}}} &=&1 + 3.75~10^{-2}(\Omo-1)~{\rm for}~\Omo+\Lambda=1  
\end{eqnarray}

The overall accuracy is similar to the accuracy obtained for the
critical Universe reference profile, for $\Omo>0.1$.  Again we
emphasize that the analytical formula must only be used between
$x_{\rm min}=0.1 \Delta_{0.05}^{1/2} f_{\rm {d_{A}}}$ and $x_{\rm
max}=0.7 \Delta_{0.05}^{1/2} f_{\rm {d_{A}}}$.

\section{Computation of the $\chi^{2}$ value}
\label{sec:chi2}

\def \xk {x_{k}} 
\def \yk {y_{k}} 
\def \sxk {\sigma_{x_{k}}} 
\def \syk {\sigma_{y_{k}}} 
\def \rk
{\rho_{k}}

Here, we describe the way we compute the $\chi^{2}$ value used to
compare the set of scaled emission measure profiles of distant
clusters to the reference profile.\\

The data consist of a set of scaled emission measure, $y_{k} =
\widetilde{EM_{k}}$, measured at the scaled radius $x_{k}$.  These
quantities are derived from the surface brightness, $SB_{k}$, at
angular radius $\theta_{k}$ (corresponding to the scaled radius
$x_{k}$) and the temperature of the specific cluster, $T_{k}$:
\begin{eqnarray}
y_{k} &\propto & SB_{k}~T_{k}^{\alpha}\\
x_{k} &\propto & T_{k}^{1/2}
\end{eqnarray}
where $\alpha$ is the slope of the \EMT relation.
The corresponding errors, $\syk$ and $\sxk$ are:
\begin{eqnarray}
   \frac{\sxk}{\xk} &= &\frac{1}{2}\ \frac{\sigma_{T_{k}} }{T_{k}}\\
   \frac{\syk^{2}}{\yk^{2}} &= &\frac{ \sigma_{SB_{k}}^{2} }{SB_{k}^{2}} +
   \alpha^{2} \frac{\sigma_{T_{k}}^{2} }{ T_{k}^{2} }
\end{eqnarray}
where $\sigma_{T_{k}}$ and $\sigma_{SB_{k}}$ are the uncertainties on
$T_{k}$ and $SB_{k}$ respectively.  The errors on $\xk$ and $\yk$ are thus
correlated through the error on the temperature.  The correlation
factor $\rho_{k}$ is:
\begin{eqnarray}
\frac{\rk \sxk \syk}{\xk \yk} &=&\frac{\alpha}{2}\
\frac{ \sigma_{T_{k}^{2}} }{T_{k}^{2}}
\label{eq:rho}
\end{eqnarray}

Let us note $Y= f(X)$ the equation of the reference curve to which this
data set is compared.  In practice, it is given in tabular form and
the data for any value of $X$ is obtained by interpolation.  The
$\chi^{2}$ expression can be found in York (\cite{york}) for the case
of correlated errors:
\begin{eqnarray}
\chi^{2}& =&\sum_{k=1}^{N} S_{k}
\end{eqnarray}
where $N$ is the number of data points and $S_{k}$ the distance of the
data point $k$ to the reference curve, which is obtained by minimizing
over $X$ the function:
\begin{eqnarray}
S_{k}(X) & = &\frac{1}{ 1-\rk^{2} }
 \left[\frac{ \left(\xk-X\right)^{2} }{\sxk^{2}} ~+ \frac{
 \left(\yk-f(X)\right)^{2} }{\syk^{2}}\right. \nonumber \\
& & ~~~~~~~~~~~\left. - 2 \rk \frac{\left(\xk-X\right)
\left(\yk-f(X)\right) }{ \sxk \syk } \right]
\label{eq:S}
\end{eqnarray}
Since the reference function is not linear, this minimization, which
actually determines the `closest' reference point, is done
numerically.  Up to this stage, we have not taken into account the
dispersion, $\sigma_{f(X)}$ at radius $X$, observed around the
reference curve.  This is done by adding quadratically this dispersion
to $\syk$.  Eq.~\ref{eq:rho} and Eq.~\ref{eq:S} remain the same, where
$\syk$ is replaced by $\sqrt{\syk^{2} + \sigma_{f(X)}^{2}}$.

\end{document}